\documentstyle[psfig]{mn}

\def\etal{{\it et al. }}

\begin{document}

\title[
Galaxy Ages and Metallicities
]
{
A Catalogue and Analysis of Local Galaxy Ages and Metallicities
}

\author[
A.~I.~Terlevich and Duncan A. Forbes
]
{
A.~I.~Terlevich$^1$ and Duncan A. Forbes$^2$\\
$^1$School of Physics and Astronomy, 
University of Birmingham, Edgbaston, Birmingham B15 2TT \\
ale@star.sr.bham.ac.uk\\
$^2$Centre for Astrophysics \& Supercomputing, Swinburne University,
Hawthorn VIC 3122, Australia\\
dforbes@swin.edu.au\\
}

\date{16th July 2001}
\pagerange{\pageref{firstpage}--\pageref{lastpage}}
\def\LaTeX{L\kern-.36em\raise.3ex\hbox{a}\kern-.15em
    T\kern-.1667em\lower.7ex\hbox{E}\kern-.125emX}

\newtheorem{theorem}{Theorem}[section]

\label{firstpage}

\maketitle

\begin{abstract}
We have assembled a catalogue of relative ages, metallicities and 
abundance ratios for
about 150 local galaxies in field, group and cluster environments. The
galaxies span morphological types from cD and ellipticals, to late
type spirals. Ages and metallicities were estimated from
high quality published spectral line indices 
using Worthey \& Ottaviani (1997) single stellar population
evolutionary models.

 The identification of galaxy age as a fourth parameter in the
fundamental plane (Forbes et al. 1998) is confirmed by our
larger sample of ages. We investigate trends between age and
metallicity, and with other physical parameters of the galaxies, such
as ellipticity, luminosity, and kinematic anisotropy.
  We demonstrate the existence of a galaxy age--metallicity
relation similar to that seen for local galactic disk stars,
whereby young galaxies have high metallicity, while old galaxies span
a large range in metallicities. 

  We also investigate the influence of environment and morphology on
the galaxy age and metallicity, especially the predictions made by
semi-analytic hierarchical clustering models (HCM). We confirm that
non-cluster ellipticals are indeed younger on average than cluster
ellipticals as predicted by the HCM models. However we also find
a trend for
the more luminous galaxies to have a higher [Mg/Fe] ratio than the
lower luminosity galaxies, which is opposite to the expectation
from HCM models.

\end{abstract}


\section{Introduction}

Elliptical galaxies were once thought to be very old ($\sim$15 Gyrs)
systems, forming in a simple monolithic collapse, and evolving
passively ever since. From colour--magnitude (e.g. Bower \etal 1992;
Ellis \etal 1997; Kodama \etal 1998), Mg--$\sigma$ (e.g. Ziegler \&
Bender 1997; Bernardi \etal 1998) and M/L studies (e.g. van Dokkum
\etal 1998) it appears that most stars in elliptical galaxies formed
at z $>$ 2. However, even though the bulk of the star formation
occurred at high redshift, the evolutionary history of ellipticals is
more complex. In terms of the star formation history, there is a
variety of evidence for more recent activity. In the Local Group, the
dwarf spheroidal galaxies reveal signs of younger stellar populations
(see review by Grebel 1997). Some nearby giant ellipticals show
H$\alpha$ line emission (Goudfrooij \etal 1994) indicative of ongoing
star formation (albeit at a low rate), E~+~A spectra indicating the
presence of a secondary starburst in the last few Gyrs (Caldwell \etal
1993; Zabludoff \etal 1996), and blueward deviations from cluster
colour--magnitude relations due to young stellar components
(e.g. Terlevich \etal 1999).  The presence of young globular clusters
also suggests recent star formation events in ellipticals
(e.g. Whitmore \etal 1997; Brown \etal 2000).  The cause of this
extended star formation history is probably driven by gaseous mergers
and accretion events. For example, Schweizer \& Seitzer (1992) showed
that ellipticals with blue galaxy colours (associated with recent star
formation) also reveal morphological disturbances (suggestive of
dynamical youth).

On the theoretical side, the `monolithic collapse scenario' has been
challenged by the idea of hierarchical clustering and merging (HCM) of
disk galaxies, and their dark matter halos, to form ellipticals
(e.g. Kauffmann \& Charlot 1998; Baugh \etal 1998). In this model,
environment plays a key role in determining the evolution of an
elliptical; cluster ellipticals are assembled at high redshift, while
their field counterparts formed much more recently. These different
evolutionary paths can be probed by examining the star formation
history.

Directly measuring the age of the stars in elliptical galaxies has
been problematic due to the well known age--metallicity degeneracy of
old stellar populations. However it is now possible, with the
combination of moderate resolution spectra and new stellar population
models to break this degeneracy, and estimate the relative ages and
metallicities of stellar populations independently (e.g. Gonz\'{a}lez
1993; Worthey 1994).  To quote Governato \etal (1999), ``These
[age--dating] methods will prove invaluable in tracing the origin of
early--type galaxies in different environments and will provide a
larger database to test theories of galaxy formation.''

In an initial study exploiting these new age estimates, we examined
the scatter about the fundamental plane (FP) for elliptical galaxies
with galaxy age (Forbes, Ponman \& Brown 1998).  We found a strong
correlation indicating that a galaxy's position relative to the FP
depends on its age, and showed that this age is consistent with the
idea that it traces the last major episode of star formation, which in
turn was presumably induced by a gaseous merger event. An analysis of
the ages and metallicities of galaxies from the small samples of
Gonzalez (1993) and Kuntschner (2000) has recently been carried out by
Trager et al. (2000)

In this paper we have set out to compile a catalogue of high quality
ages, metallicities and abundances for (mostly) early type galaxies from the
literature.  These ages are based on central absorption line indices
calibrated to the Lick system.  Such a catalogue will have many
applications. For example, after galaxies have been age--dated we may
begin to explore an `evolutionary sequence' of elliptical galaxies
tracking their energetic, chemical, structural and dynamic properties
as they age. Another use is to test various predictions of the HCM
models as mentioned above. We hope that this catalogue will be of use
to many researchers involved in galaxy evolutionary studies. Due to
their size, Tables 2 and 4 are available electronically from
author DF or from
http://astronomy.swin.edu.au/staff/dforbes/agecat.html

In section \ref{sec:LICK} we briefly explain how the spectroscopic
age--dating technique used in the paper works, and in section
\ref{sec:caveats}, we outline some of the limitations of this
method. This paper is based on data mined from many sources. Section
\ref{sec:Samples} details each source, and any additional steps
necessary to merge them into a homogeneous catalogue.
We present this catalogue in section \ref{sec:cat}. In section
\ref{sec:results} we investigate how the ages and metallicities of the
galaxies from the catalogue correlate with the physical parameters of
the galaxies, such as luminosity, anisotropy parameter, morphology and
environment. We comment on our findings in light of predictions from
models.

\section{The Lick System And Galaxy Ages}
\label{sec:LICK}
The Lick system is perhaps the most widely used definition of
absorption line indices for old stellar populations. It is based on
spectra obtained with the Image Dissector Scanner (IDS) on the Lick 3m
telescope, by S. Faber and collaborators (e.g. Faber \& Jackson 1976;
Burstein \etal 1984). These spectra cover the wavelength range
$\sim$4000--6400\AA~ with $\sim$9\AA~ resolution. Further details can
be found in Worthey \etal (1994), Worthey \& Ottaviani (1997) and
Trager \etal (1998).

As well as creating an extensive library of galaxy, globular cluster
and stellar spectra, new stellar population models were developed to
accompany the Lick indices (Worthey 1994). The relative age and
metallicity sensitivity of each line index was quantified, thus
allowing the well known age--metallicity degeneracy of old stellar
populations to be broken. In this paper, we have chosen to use the
H$\beta$ line index and the combination index [MgFe] as these have
good age and metallicity sensitivity respectively, and are available
for many galaxies.

The H$\beta$ line index is defined between 4847.875 and 4876.625\AA~
with continua either side. It has the advantage of being sensitive to
stellar age and a relatively strong line at a wavelength where most
CCDs have a good quantum efficiency. One serious disadvantage of
H$\beta$ is that it suffers from nebular emission in some galaxies,
which `fills in' the absorption line (Gonz\'{a}lez 1993). The bluer
Balmer series (e.g. H$\gamma$ and H$\delta$) suffer far less from this
problem (Osterbrock 1989) but these indices are slightly less age
dependent (Worthey \& Ottaviani 1997) and are more difficult to
measure accurately than the H$\beta$ index, so have not yet been
measured for a large sample of galaxies.

In luminous elliptical galaxies, Mg appears to be overabundant
compared to Fe (e.g. Worthey et al. 1992; Davies et al. 1993). The
fitting functions used in the Worthey models are derived from the Lick
stellar library, which contain stars in the solar neighbourhood and
are thus of mainly solar abundance ratios (McWilliam 1997). If the
galaxies being studied have non solar abundance ratios, then the ages
and metallicities will not necessarily be accurate. The combination
index [MgFe] was defined by Gonz\'alez to be an average of the Mg
and Fe, and is thus a better tracer of metallicity than either Mg
or Fe alone. It is calculated from (${\rm Mg}b~ \times <$Fe$>)^{1/2}$,
where $<$Fe$>$ is the average of the Fe52 and Fe53 line indices. The
individual line indices are defined as 5160.125 to 5192.625\AA~ for
Mg$b$, 5245.650 to 5285.650\AA~ for Fe52 and 5312.125 to 5352.125\AA~
for Fe53. Due to the overabundance issues, and the widespread
availability of [MgFe] index values in the literature, we have used it
as the metallicity sensitive index for this work.
We note that [NI] 5199\AA~ emission may contribute to the
red continuum band of Mg$b$ in some galaxies (Goudfrooij \& Emsellem
1996). This would tend to make galaxies appear slightly 
younger than they really are. This is unlikely to be a strong
effect in our sample as we have removed galaxies with evidence
for emission lines (see Table 1). Furthermore the Mg$b$ index is
combined with two Fe indices which do not suffer from emission.

Although H$\beta$ line indices on the Lick system now exist for over
$\sim$ 500 galaxies, not all were obtained at sufficient S/N to derive
accurate ages. Our aim here is to compile a relatively homogeneous
sample of high quality H$\beta$ line indices. 
{\it From the literature we have included
galaxies that have EW$($H$\beta)$ measurement errors of $\le \pm
0.1$\AA~}. Assuming no errors in [MgFe], this corresponds to an
uncertainty in age of about $\pm 20\%$, and an uncertainty of $\pm
0.2$ dex in [Fe/H] for a $\sim 4$ Gyrs old galaxy. The mean measurement
errors in our final sample are 0.6\AA~for both H$\beta$ and [MgFe]
(see section \ref{sec:corr} for a discussion on the effects of
correlated errors) . A number of published studies meet our H$\beta$
quality criteria, and these are discussed briefly below. Unfortunately
several other studies do not, in general, meet this criteria
(e.g. Gorgas \etal 1990; J$\o$rgensen 1997, 1999; Trager \etal
1998). The H$\beta$ errors in these studies are typically $\pm$
0.3\AA~ which effectively rules out individual age determinations from
such data.

Our basic approach is to collect central (i.e. small aperture)
H$\beta$ and [MgFe] line indices from the literature, and via
interpolation of the Worthey single stellar population (SSP)
evolutionary tracks (Worthey 1994; Worthey \& Ottaviani 1997), derive
age and [Fe/H] metallicity values. We also calculate the
Mg$b$/$<$Fe$>$ ratio as an indicatior of the [Mg/Fe] abundance
ratio. Before discussing the samples, we remind the reader of several
caveats that should be born in mind before interpretation of any
results presented in this paper.

\section{Caveats}
\label{sec:caveats}

\subsection{Models}

In this paper we have chosen to use the models of Worthey \& Ottaviani
(1997) to translate measured line indices into age and metallicity
estimates. The results are model dependent. Several alternative models
are available and include: Buzzoni \etal (1992, 1994), Bruzual \&
Charlot (1993), Fritze-von Alvensleben \& Burkert (1995) and Vazdekis
et~al. (1996). Recently Maraston, Greggio \& Thomas (2001) have
studied the differences between the model predictions of Worthey \&
Ottaviani (1997) and Buzzoni \etal (1992, 1994). They find, in
particular, that the values assumed for the Horizontal Branch
evolutionary mass and hence T$_{eff}$ have a large effect on the
predicted Fe and H$\beta$ line strengths. This leads to different
derived ages. For example, at solar or super-solar metallicities
Worthey models predict younger ages than Buzzoni \etal of $\sim$ 1--2
Gyrs. However the situation is worse for metal--poor galaxies, with
age differences up to 4 Gyrs being possible. Fortunately most of the
galaxies in our sample have central metallicities that are at least
solar according to their [MgFe] index.

\begin{figure}
\centerline{\psfig{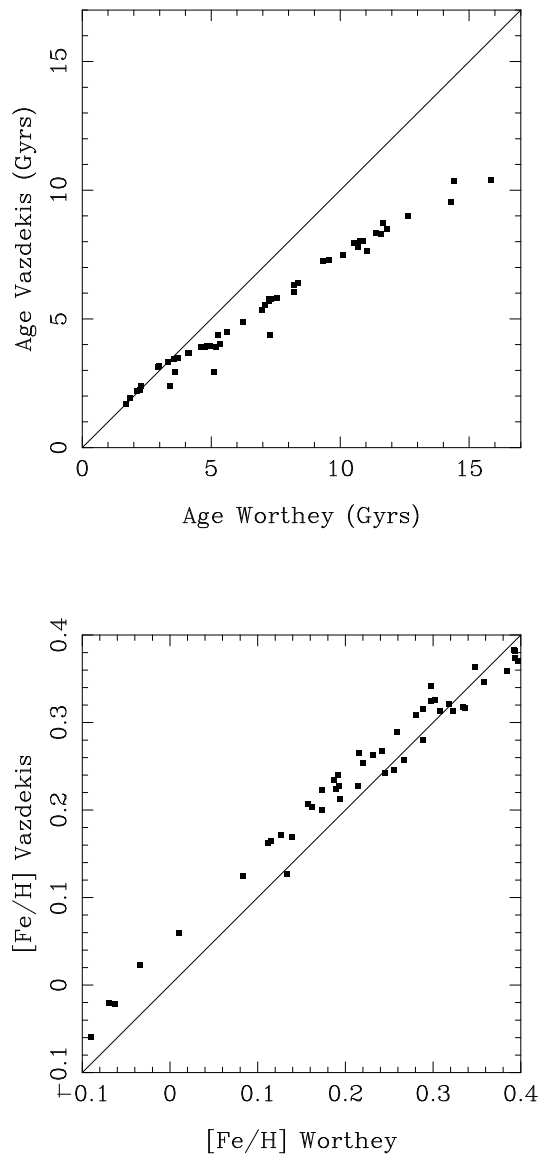}}
\caption{ A comparison of the ages and [Fe/H] determined for some of
our sample when using the Worthey models used throughout the rest of
this paper, and the models of Vazdekis et al. (1996). The [Fe/H]
estimated from both models are in good agreement, however the ages
estimated from the Vazdekis models are systematically younger than
those given by the Worthey models. This is an indication of how these
models should not be used to estimate absolute ages for these systems,
however the ranking of the galaxies is preserved, such that it is
possible to order the galaxies from young to old. This is sufficient
for the analysis which follows in this paper. }
\label{fig:VWcomp}
\end{figure}

 To check the degree to which using a different model would affect our
results, we have estimated ages and metallicities for some of the
galaxies in our sample using both the Worthey models used throughout
this paper, and those of Vazdekis et al. (1996). The Vazdekis models
use a superset of the stars used by Worthey (1994) to calculate the
fitting functions that produce the Lick indices from the models,
however they use different isochrones to determine the SSP HR
diagram. Figure \ref{fig:VWcomp} shows that while the [Fe/H] values
agree well between the models, the Vazdekis models systematically
predict younger ages than the Worthey models do. This is not a problem
per se, as both models give the same ranking of the galaxies in terms
of age, but it should be noted that the ages we quote in this paper
should be considered as {\it relative} to each other, and not absolute.

\subsection{Non--solar Abundances}

It has been known for some time that the most luminous ellipticals
have non--solar abundance ratios, e.g. [Mg/Fe] $\sim$ +0.3 (O'Connell
1976; Peletier 1989; Worthey, Faber \& Gonz\'{a}lez 1992; Carollo
\etal 1993; Vazdekis \etal 1997). Unfortunately spectra of stars with
enhanced $\alpha$ elements are not generally available. So using Fe or
Mg alone will result in significantly different age estimates. The
combination index [MgFe] to some extent overcomes this problem by
providing a `mean' metallicity indicator, although there is still a
slight tendency to underestimate the age of the most massive
ellipticals.

As well as estimating ages and [Fe/H] metallicities for galaxies, we
use the ratio of Mg$b$ divided by $<$Fe$>$ as a proxy for the [Mg/Fe]
abundance ratio (see discussion by Matteucci \etal 1998).

\subsection{Aperture Effects}

Studies of line index radial gradients have generally found most early
type galaxies to have a small variation from young metal--rich centers
to older and more metal--poor stars in the outer parts (Davies \etal
1993; Carollo \etal 1993, Gonzalez 1993). Thus different apertures may
affect the derived mean age and metallicity. In this paper, we have
calculated mean line indices within $R_e/8$ for the samples with
radial gradient information and used the $R_e/8$ values quoted by
Gonzalez (1993). This is achieved by weighting the profile by a
canonical $R^{1/4}$ surface brightness law:\\
\begin{equation}
\label{eqn:linegrad}
\bar{I}_{R_e\over8} = \frac {\sum_{i=1}^{n} I(R_i) L(R_i)} {\sum_{i=1}^{n} L(R_i) }
\end{equation}
where $L(R)$ is the $R^{1/4}$ profile surface brightness at
a radius $R$, and $I(R)$ is the value of the spectral line index at
radius $R$. The $R^{1/4}$ profile is normalised such that $I_e=1$:
\begin{displaymath}
\label{eqn:sb}
L(R) = exp \left[-7.67\left(\frac{R}{R_e}\right)^{1/4}-1\right]
\end{displaymath}

To calculate $\bar{I}$ within an aperture of $R_e/8$, we use a value
of $n$ in equation \ref{eqn:linegrad} such that
\begin{displaymath}
R_n<R_e/8<R_{n+1}
\end{displaymath}

Some of the samples used here do not have radial gradient information
available. In the case of data from Carollo \etal (1993) and
Kuntschner (1998) we use their fixed aperture measurements at 3$^{''}$
and 2.8$^{''}$ respectively. Mehlert \etal (1997) uses $R_e/10$.
Although aperture effects will be present (e.g. apertures larger than
$R_e/8$ may tend to increase the derived age and reduce the
metallicity) they tend not to be large since observed radial gradients
are fairly shallow.

\begin{figure}
\centerline{\psfig{figure=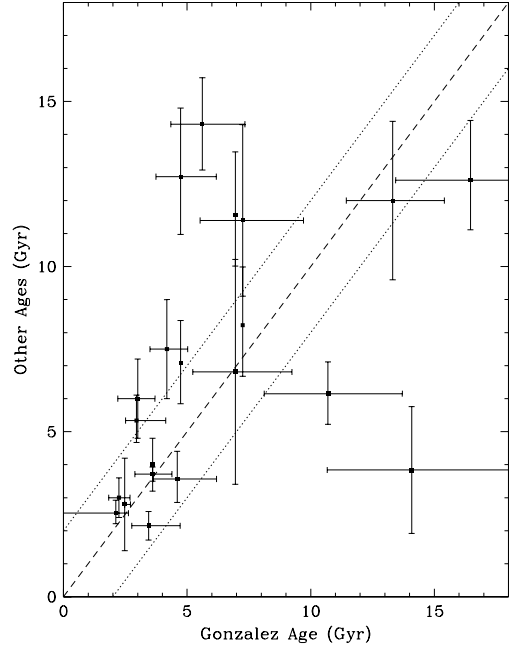,angle=0,width=200pt}}
\caption{ Comparison between individual ages derived from different
samples for the same galaxy, and the age for these galaxies from the
Gonz\'alez (1993) sample. The dotted lines show a $\pm 2$ Gyrs spread
about the unity (dashed) line. There is reasonable agreement over a
wide range of galaxy ages.
}
\label{fig:comp_meanage}
\end{figure}

We can begin to probe the importance of aperture effects in our sample
by comparing different age estimates for the same galaxies.  In
Figure \ref{fig:comp_meanage} we show the age estimates for galaxies
that are in common with the Gonzalez (1993) sample (a sample that uses
$R_e/8$). The figure shows a one-to-one unity line and $\pm$ 2 Gyrs
about this line. The bulk of the galaxies have ages within 2 Gyrs of
that derived from the Gonzalez data, over the full age range.

There are however notable exceptions with large age differences.  They
are NGC 2778 (giving the Gonzalez derived age first) 4.7 vs 12.7 Gyrs
from Fisher \etal (1995), NGC 3608 5.6 vs 14.3 from Halliday \etal
(1998) and NGC 7619 14.1 vs 3.8 Gyrs from Fisher \etal (1995). It is
impossible with just two age estimates to decide which better
represents the age of the galaxy. We note however, that in these three
cases aperture effects are probably not the cause as the Fisher \etal
and Halliday \etal line indices were all corrected to be at $R_e/8$,
same as the Gonzalez (1993) data.  In the final catalogue (Table 2) we
give the simple mean of all different age estimates.

\subsection{Emission}
\label{sec:emission}

 We have chosen to use the H$\beta$ line index as the dominant age
indicator, since this is the most widely available. It has the
advantage of being a relatively strong absorption line but can be
affected by nebular emission which can rapidly fill the absorption
feature, making the galaxy appear older than it really is.
Gonz\'{a}lez (1993) made an empirical correction for H$\beta$ emission
based on the [OIII]~5007\AA~ equivalent width, i.e. H$\beta_{new}$ =
H$\beta_{old}$ + EW([OIII]) $\times$ 0.7.  For an individual galaxy this
correction may not be entirely correct, but on average it makes the
H$\beta$ line index a more accurate age indicator.  Some of the
galaxies in the samples of Davies \etal (1993), Carollo \etal (1993),
Vazdekis \etal (1997) and Halliday (1998) are in common with those of
Gonz\'{a}lez (1993). Furthermore we have derived H$\beta$ line indices
in an $R_e/8$ aperture for these four samples and for Gonz\'{a}lez
(1993).  We thus correct all common galaxies using the Gonz\'{a}lez
(1993) [OIII] measurements. Both Fisher samples (Fisher \etal 1995,
1996) have been H$\beta$ emission corrected using their own
measurements of [OIII].  For all samples, galaxies with strong
emission have been excluded from this analysis and are listed in Table
1. The ages derived by Kuntschner (1998) (see sec. \ref{sec:KD_data})
used the H$\gamma$ Balmer line index, which is far less sensitive to
nebular emission than H$\beta$, but to be consistent with the other
samples, we have used ages derived from their H$\beta$ and [MgFe] measurements.

\subsection{Luminosity Weighting}

The galaxy ages and metallicities derived from absorption lines and
stellar population models assume that a single stellar population is
present. If more than one population of stars is present within the
measurement aperture (which is the most likely case), then the derived
ages will be dominated by the younger stellar population. This is
because the strength of the line indices reflect the luminosity of the
stellar population that produces them, and that young stars are
particularly luminous at blue wavelengths. Thus even a small
percentage, by mass, of young stars mixed with old stars can dominate
the resulting age estimate.

\begin{figure}
\centerline{\psfig{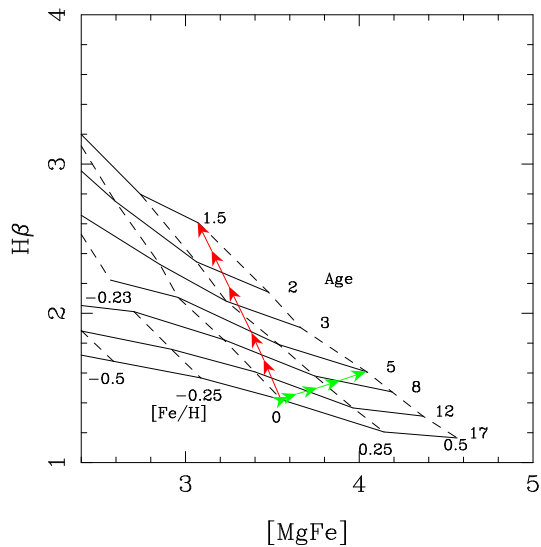}}
\caption{ Absorption line model grid (from Worthey \& Ottaviani (1997)
showing the effects of a young stellar population combined with an old
(17 Gyrs) solar metallicity population. The arrows represent 5\%, 10\%,
25\%, 50\% and 100\% contributions by mass of a super solar, young
(1.5 or 5 Gyrs) populations. These `mixing curves' encompass the bulk
of the galaxy data.  }
\label{fig:MSP_effect}
\end{figure}

The effect of young stars mixed with an old stellar population is
illustrated in Figure \ref{fig:MSP_effect}. Here we start with a solar
metallicity, old (age = 17 Gyrs) stellar population. We then add
increments of a metal--rich ([Fe/H] =0.5), young population (1.5 and 5
Gyrs are shown) of various strengths. Even a small fraction of young
stars results in an inferred age that is much less than 17 Gyrs.

\subsection{Correlated Errors}
\label{sec:corr}

While the errors on the measured indices are independent, the model
tracks are not orthogonal (e.g. Figure \ref{fig:MSP_effect}). This
produces non orthogonal errors in age and metallicity which can lead
to spurious correlations if not understood properly. Additionally, the
degree of non orthogonality of the age and metallicity tracks changes
with position on the grid. For this reason we have used Monte Carlo
methods to model the effects of these correlations on 'ideal' initial
data distributions.

  The model works by assuming an initial linear distribution between
age and metallicity. A large sample of `galaxies' is drawn at random
from this distribution. We then use the Worthey models to assign
[MgFe] and H$\beta$ index values to these galaxies, and at this stage,
we add the Gaussian errors to the Lick indices. The errors are drawn
at random from a distribution whose $\sigma$ corresponds to the mean
quoted error for that index in the real data. Once we have a set of
model data with errors, we process them in the same way as the real
galaxies, to produce a model age-metallicity catalogue. This can
be compared to the assumed initial age-metallicity distribution, using a
2d Kolmogorov-Smirnov (KS) test, or converted into contours which
contain $99.56\%$ of the model galaxies, comparable to a $3\sigma$
error ellipse. In section \ref{sec:TZreln} we compare the data to
models for a single age initial population, a single metallicity
initial population and a population where the metallicity ([Fe/H])
depends linearly on age.

\section{The Samples}
\label{sec:Samples}

\begin{table*}
\caption{Galaxies excluded from the catalogue. We have
excluded galaxies due to suspected H$\beta$ emission filling, too
large a quoted error on the  H$\beta$ index, or a lack of either
H$\beta$ or Fe index measurements. See text for details.}
\centerline{
\begin{tabular}{lllllll} 
Sample & Galaxy & Reason & Galaxy & Reason & Galaxy & Reason\\
\hline
Davies & NGC 315 & Emission & NGC 741 & Emission 
& NGC 4486 & Emission\\
Davies & NGC 4636 & Emission & NGC 4839 & No Fe obs. & & \\
Carollo & ESO208-21 & Emission & ESO323-16 & Emission
& ESO381-29 & Emission\\
Carollo & IC 1459 & H$\beta$ error & IC 2006 & H$\beta$ error & 
IC 2035 & H$\beta$ error\\
Carollo & IC 3370 & H$\beta$ error & IC 4889 & H$\beta$ error
& IC 4943 & Emission\\
Carollo & NGC 1052 & H$\beta$ error 
& NGC 1298 & Emission & NGC 1947 & Emission\\
Carollo & NGC 2502 & H$\beta$ error & NGC 2663 & H$\beta$ error
& NGC 2974 & H$\beta$ error\\
Carollo & NGC 3108 & H$\beta$ error
& NGC 3136 & Emission & NGC 3136B & Emission\\
Carollo & NGC 3226 & Emission & NGC 3250 & H$\beta$ error
& NGC 3260 & Emission\\
Carollo & NGC 3557 & Emission
& NGC 4374 & H$\beta$ error & NGC 4684 & Emission\\
Carollo & NGC 4696 & Emission & NGC 4832 & Emission
& NGC 5011 & H$\beta$ error\\
Carollo & NGC 5044 & Emission 
& NGC 5077 & Emission & NGC 5090 & H$\beta$ error\\
Carollo & NGC 5266 & Emission & NGC 5796 & H$\beta$ error
& NGC 5846 & H$\beta$ error\\
Carollo & NGC 5903 & Emission
& NGC 6849 & Emission & NGC 7097 & H$\beta$ error\\
Carollo & NGC 7200 & H$\beta$ error & & & &\\
Fisher95 & A 496 & No H$\beta$ obs. & NGC 5846 & Emission & 
NGC 6166 & Emission\\
Fisher95 & NGC 7720 & Emission \\
Fisher96 & NGC 2560 & H$\beta$ error & NGC 3998 & H$\beta$ error  & 
NGC 4550 & No Fe obs.\\
Mehlert & NGC 4850 & Emission & & & &\\
Vazdekis & NGC 4594 & Emission & & & &\\
Longhetti & E244-G12 & Emission & E289-G15 & Emission 
& E386-G04 & Emission\\
Longhetti & IC 4823 & Emission & IC 5063 & Emission & NGC 7135 
& Emission\\
\hline
\end{tabular}
}
\label{tab:Exclude}
\end{table*}

\subsection{Davies \etal}

The earliest study in our compilation is that of Davies, Sadler \&
Peletier (1993). Their spectra of 13 luminous elliptical galaxies were
taken on the KPNO 4m telescope using two different spectrographs. Here
we are interested in the central line indices which come mostly from
the RC spectrograph with a resolution of 6\AA~. For our purposes we
will use their radial gradient measurements of H$\beta$, Fe52 and Fe53
to calculate mean values within $R_e/8$, using equation 1 above. As
the Davies \etal study does not include ${\rm Mg}b$ measurements, we do
not include it in our analysis. We do however provide our derived ages
and metallicities for Davies \etal in Table 2.

\subsection{Carollo \etal}

After several observing runs on the ESO 3.6m, Carollo \etal (1993)
obtained spectra for 42 early type galaxies. The spectral resolution
was about 9\AA~. They averaged the inner 3$^{''}$ to obtain the mean
central values for each line index. Many galaxies had to be excluded
due to strong emission, or large H$\beta$ errors; these galaxies are
listed in Table 1.  They did not measure Mg$b$ or Fe53, so we use the
Worthey models for the Fe52 and H$\beta$ indices to derive ages and
metallicities, but do not include them in our analysis. We do however
provide our derived ages and metallicities for Carollo \etal in Table
2.

\subsection{Gonz\'{a}lez}

Using the Lick 3m telescope, Gonz\'{a}lez (1993) obtained spectra of
41 galaxies with a spectral resolution of $\sim$3.5 \AA~.  The
H$\beta$ line index was corrected for emission by scaling from the
strength of [OIII] emission.  Here we have chosen to use his $R_e/8$
aperture size. Most of the galaxies in his sample are located in low
density environments, although five are in the Virgo cluster.

\subsection{Fisher \etal}

In two papers, Fisher, Franx \& Illingworth (1995, 1996) investigated
the line strength gradients in a large number of early type
galaxies. They obtained spectra at the Lick 3m with a resolution of
$\sim$3\AA~ and at the KPNO 4m telescope with spectral resolution
varying from 6.5 to 11\AA~. The combined sample includes 20 S0
galaxies, 2 ellipticals, 8 brightest cluster galaxies and 1 compact
elliptical (M32).  Radial gradient information was obtained for
several line indices including H$\beta$, Mg$b$, Fe52 and Fe53. We have
weighted each individual line measurement assuming an R$^{1/4}$
profile and derived an average line index within $R_e/8$ using
equation \ref{eqn:linegrad}.

\subsection{Mehlert \etal}

\begin{figure}
\centerline{\psfig{figure=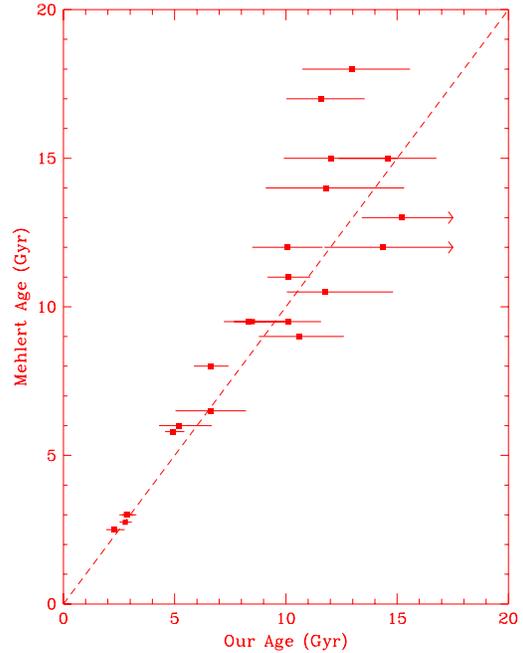,width=200pt}}
\caption{ Comparison between ages derived by Mehlert (1999, private
communication) and by us for Coma cluster galaxies. There is good
agreement, except perhaps for the very oldest galaxies.  }
\label{fig:comp_Mehlert}
\end{figure}

Using several different telescopes, Mehlert \etal (1997) obtained
spectra for 35 early type galaxies in the Coma cluster. Of these, 14
represent a complete sample down to M$_B$ = --21.6. They derived
radial gradient information for the Mg, Fe and H$\beta$ line
indices. We have used their $R_e/10$ aperture measurements. NGC 4839
has been corrected for H$\beta$ emission using the [OIII] measurements
from Fisher \etal (1995), and NGC 4850 has been excluded due to strong
H$\beta$ emission. We have used the radial line index measurements to
exclude portions of galaxies showing signs of emission.  A comparison
between our derived ages, and those of Mehlert (1999, private
communication) are shown in Figure \ref{fig:comp_Mehlert}. There is
generally good agreement, except perhaps for the very oldest galaxies.

\subsection{Vazdekis \etal}

Three well known early type galaxies (i.e. NGC 3379, 4472 and 4594)
were observed by Vazdekis \etal (1997) using the 4.2m WHT. They
obtained radial gradient information for several line indices at
$\sim$5\AA~ resolution.  After excluding NGC 4594 due to strong
emission, we have calculated mean indices within $R_e/8$ for the
remaining two galaxies using equation~\ref{eqn:linegrad}.

\subsection{Longhetti \etal}

Longhetti \etal (1998) studied a sample of early type galaxies in low
density environments. In particular they obtained spectra, at a
resolution of 2.1\AA~, for 21 galaxies with shells and 32 in pairs.
Longhetti \etal identified those galaxies with signs of H$\beta$
emission. These have been excluded from our analysis and are listed in
Table 1.  We have velocity--corrected their published data using the
formulation quoted in Longhetti \etal (1998).

\subsection{Halliday}

In her thesis, Halliday (1998) presents MMT observations of 14 low
luminosity early type galaxies. Eight are located in the Virgo cluster
and six in lower density environments. The spectra are high S/N with a
resolution of 1.5\AA~. From the radial gradient information we have
derived mean line indices within $R_e/8$ using equation
\ref{eqn:linegrad}. Where possible galaxies have been corrected for
H$\beta$ emission using [OIII] from Gonz\'{a}lez (1993), otherwise we
use the H$\beta$ radial profile information from Halliday (1998) to
determine and exclude portions of the galaxy most affected by
emission, as was done with the Mehlert \etal (1997) data.

\subsection{Kuntschner}
\label{sec:KD_data}

Kuntschner (1998; 2000) obtained spectra for a complete sample of
early type galaxies in the Fornax cluster to M$_B$ = --17.1.  This
included the cD galaxy NGC 1399 and the peculiar galaxy NGC 1316 (also
known as Fornax A).  Line indices were measured for an effective
aperture of 2.8$^{''}$ (230 pc). For their typical galaxy this is
about $R_e/10$.  In addition to H$\beta$, Mg, Fe, they measured
H$\gamma$ and C4668.  Using their H$\beta$ values (corrected for
emission where possible) and [MgFe] indices, as given in Kuntschner
(2000), we have derived ages and metallicities for their sample.

\begin{figure}
\centerline{\psfig{figure=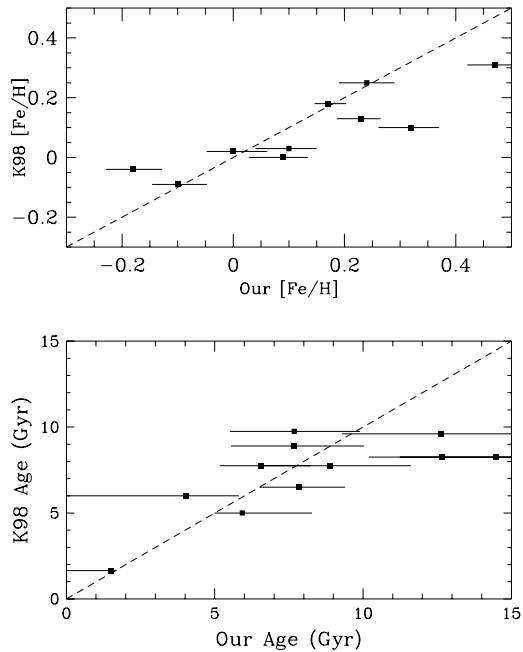,width=200pt}}
\caption{ Comparison between ages and metallicities, for Fornax
cluster galaxies, derived from H$\beta$--[MgFe] by us and from
H$\gamma$--C4668 by Kuntschner (1998). Both methods give similar
[Fe/H] values over all metallicities, and reasonable agreement in ages
for ages, less than about 12 Gyrs. In the final catalogue we give the
H$\beta$--[MgFe] derived values which we use in the analysis.  }
\label{fig:comp_KD}
\end{figure}

Kuntschner (1998; 2000) derived ages and metallicities from H$\gamma$,
C4668 and Fe3 (defined in Kuntschner 2000). The H$\gamma$ line index
has the advantage of being less affected by nebular emission
(Osterbrock 1989) than H$\beta$. In Figure \ref{fig:comp_KD} we show a
comparison of the ages and metallicities, of their Fornax galaxies,
derived from H$\beta$--[MgFe] by us and from H$\gamma$--C4668 by them
(Kuntschner 1998). The figure shows that either method gives similar
[Fe/H] values over a large metallicity range, and reasonable agreement
in age, for ages less than about 12 Gyrs. We suspect that weak
emission is filling the H$\beta$ line, and causing us to overestimate
the age of these few galaxies. To remain consistent with the data from the
other authors, we use the H$\beta$--[MgFe] ages for
the analysis in this paper.

\subsection{Vazdekis \& Arimoto}

Vazdekis \& Arimoto (1999) have defined a new line index based around
H$\gamma$. It has the advantage of being very insensitive to
metallicity, while providing a robust age for old stellar
populations. Using data from Jones (1997) and Vazdekis (1996) they
derive ages for six early type galaxies. The ages for these galaxies
are taken directly from their work.

\subsection{Goudfrooij \etal }

Recently Goudfrooij \etal (1999) published an initial paper on line
strengths in 16 edge--on S0 and spiral galaxies. They placed the
spectrograph slit along the minor axis of the galaxy bulge and
carefully tried to exclude any disk contribution to the measured bulge
line indices. The spectra were obtained with the ESO NTT 3.6m and the
2.5m INT on La Palma. Their typical resolution was about 2\AA~. They
kindly supplied us with their central line indices on the Lick system,
from which we have derived ages and metallicities.

\section{The Catalogue}
\label{sec:cat}

\begin{figure}
\centerline{\psfig{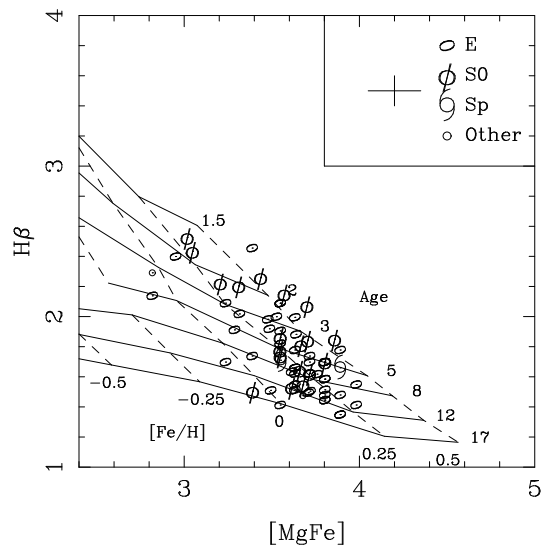}}
\caption{ Absorption line model grid showing the catalogue galaxies
with H$\beta$ and [MgFe] measurements. Symbols are coded by Hubble
type (cD and cE galaxies are classified as `other'). Solid lines
represent constant age (isochrones) and dashed lines constant
metallicity (isofers). A typical error bar is also shown.  }
\label{fig:Worthey_gals}
\end{figure}

To estimate ages and metallicities for the galaxies, we interpolate
Worthey SSP evolutionary tracks. Figure \ref{fig:Worthey_gals} shows
the distribution of galaxies from samples with H$\beta$ and [MgFe]
indices, superposed upon constant age model tracks (solid lines) and
constant metallicity model tracks (dashed lines). This figure
illustrates many of the caveats outlined in section
\ref{sec:caveats}, i.e. 

\noindent
{\it i)} The constant age and metallicity tracks
are not orthogonal, resulting in correlated errors in age and metallicity. 

\noindent
{\it ii)} Not all galaxies lie within the
parameter space covered by the models. Part of this could simply be
due to observational errors, but in part it will also be due to
incompleteness in the models, which are still evolving
rapidly. 

\noindent
{\it iii)} The constant age tracks for older galaxies are much
closer together than for young galaxies, making the errors in the ages
of older galaxies much larger than those for the younger ones. This
can clearly be seen in Figure \ref{fig:Worthey_gals}.

\begin{table}
\caption{Galaxy age, metallicity and abundance estimates. }
\begin{tabular}{l c c c c c} 
Galaxy & Age & [Fe/H] & [Mg/Fe] & Source & Quality\\
\hline

   NGC0221 &     3.8 &   -0.04 &   0.04 &      12 & 1\\
   NGC0224 &     5.1 &    0.42 &    0.21 &       5 & 1\\
   NGC0315 &     4.9 &    0.36 &    0.26 &       5 & 1\\
   NGC0507 &     6.2 &    0.21 &    0.24 &       5 & 1\\
   NGC0522 &     4.6 &   -0.06 &      -- &       5 & 1\\
   NGC0547 &     7.6 &    0.30 &    0.27 &       5 & 1\\
   NGC0584 &     2.1 &    $>$0.5 &    0.19 &       5 & 1\\
   NGC0636 &     3.6 &    0.38 &    0.15 &       5 & 1\\
   NGC0678 &     4.8 &    0.45 &      -- &       5 & 1\\
   NGC0720 &     3.4 &    $>$0.5 &    0.27 &       5 & 1\\
   NGC0813 &      -- &    $>$0.5 &    0.17 &       9 & 3\\
   NGC0821 &     7.2 &    0.22 &    0.25 &       5 & 1\\
   NGC0891 &      11 &   -0.47 &      -- &       5 & 1\\
   NGC0936 &      16 &   -0.01 &    0.19 &       4 & 1\\
   NGC0973 &     6.2 &    0.06 &      -- &       4 & 1\\
   NGC1032 &    3.1 &    $>$0.5 &      -- & 4 & 2\\
   NGC1184 &     5.9 &    0.22 &      -- &       4  & 1\\
   NGC1209 &      15 &   -0.01 &      -- & 1 & 3\\
   NGC1210 &     $>$17 &  -0.21 &    0.16 & 9 & 3\\
    NGC1316 &     3.4 &    0.25 &    0.15 &      12 & 1\\
    NGC1336 &     12.7 &   -0.1 &    0.22 &       8 & 1\\
    NGC1339 &     -- &    -- &    0.24 &       8 & 3\\
    NGC1351 &     12.6 &    0.1 &      -- &       8 & 1\\
    NGC1373 &     7.7 &    0.00 &    0.17 &       8 & 1\\
    NGC1374 &     7.7 &    0.32 &    0.23 &       8 & 1\\
    NGC1375 &     1.5 &    0.24 &  0.00 &       8 & 1\\
    NGC1379 &     8.9 &    0.09 &    0.23 &       8 & 1\\
    NGC1380 &     -- &    -- &    0.18 &       8 & 3\\
   NGC1380A &     -- &    -- &   0.02 &       8 & 3\\
    NGC1381 &     6.6 &    0.23 &    0.18 &       8 & 1\\
    NGC1399 &     -- &    $>$0.5 &    0.26 &       8 & 3\\
    NGC1404 &     5.9 &    0.47 &    0.19 &       8 & 1\\
    NGC1419 &     14.5 &   -0.18 &    0.19 &       8 & 1\\
    NGC1427 &     7.9 &    0.17 &    0.18 &       8 & 1\\
   NGC1453 &     7.6 &    0.30 &    0.23 &       5 & 1\\
   NGC1461 &     6.8 &    0.29 &    0.16 &       4 & 1\\
   NGC1549 &     7.6 &    0.15 &    0.21 &       9 & 1\\
   NGC1553 &     $>$17 &   0.03 &    0.18 & 9 & 3\\
   NGC1571 &     8.9 &    0.13 &    0.18 &       9 & 1\\
   NGC1600 &     7.3 &    0.41 &    0.25 &      5 & 1\\
   NGC1700 &     2.3 &    $>$0.5 &    0.13 &      12 & 1\\
   NGC2300 &     5.0 &    0.39 &    0.25 &       5 & 1\\
   NGC2329 &     $>$17 &      -- &      -- &       3 & 3\\
   NGC2778 &     8 &    0.25 &    0.21 &      12 & 2 \\
   NGC2832 &      12 &    0.20 &      -- &       3 & 1\\
   NGC2865 &    $<$1.5 &      -- &    0.16 &       9 & 1\\
   NGC2945 &     $>$17 &      -- &     0.3 &       9 & 3\\
   NGC3051 &     $>$17 &      -- &    0.27 &       9 & 3\\
   NGC3078 &      14 &    0.01 &      -- &       1 & 3\\
   NGC3115 &      -- &    $>$0.5 &    0.14 &       4 & 3\\
   NGC3289 &    $<$1.5 &      -- &    0.13 &       9 & 1\\
   NGC3377 &     4.1 &    0.20 &    0.22 &      12 & 1\\
   NGC3379 &     9.3 &    0.16 &    0.24 &      12 & 1\\
   NGC3384 &      -- &    $>$0.5 &     0.1 &       4 & 3\\
   NGC3412 &     1.9 &    0.42 &   0.09 &       4 & 1\\
   NGC3585 &     3.1 &    $>$0.5 &    0.14 & 4 & 2\\
   NGC3605 &     5.8 &    0.22 &   0.09 &      12 & 1\\
   NGC3607 &     3.6 &    $>$0.5 &    0.16 & 4 & 2\\
   NGC3608 &      10 &    0.16 &    0.22 &      12 & 2\\
   NGC3818 &      5.0 &    0.39 &    0.23 &       5 & 1\\
   NGC3941 &      -- &    $>$0.5 &    0.11 &       4 & 3\\
   NGC4036 &      -- &    $>$0.5 &    0.24 &       4 & 3\\
   NGC4073 &     7.5 &    0.35 &      -- &       3 & 1\\

\hline

\end{tabular}
\end{table}

\begin{table}
\begin{tabular}{l c c c c c} 
\hline

Galaxy & Age & [Fe/H] & [Mg/Fe] & Source & Quality\\
\hline

   NGC4105 &     $>$17 &      -- &    0.28 &       9 & 3\\
   NGC4106 &      14 &    0.04 &    0.23 &       9 & 1\\
   NGC4239 &     5.5 &      -- &      -- &      11 & 1\\
   NGC4251 &     1.9 &    0.48 &    0.09 &       4 & 1\\
    NGC4261 &     14.4 &    0.19 &    0.28 &      5 & 1\\
   NGC4278 &     10.7 &    0.14 &    0.17 &      5 & 1\\
   NGC4339 &     7.9 &    0.12 &    0.18 &       7 & 1\\
   NGC4350 &     9.3 &    0.30 &    0.16 &       4 & 1\\
   NGC4374 &      11.8 &    0.12 &    0.28 &      5 & 1\\
   NGC4382 &     1.6 &    0.44 &   0.08 &       4 & 1\\
   NGC4387 &      13 &   -0.04 &    0.17 &      12 & 1\\
   NGC4458 &      16 &   -0.30 &    0.23 &       7 & 1\\
   NGC4464 &     $>$17 &      -- &    0.23 &       7 & 3\\
   NGC4468 &     2.7 &    0.08 &   0.05 &       7 & 1\\
   NGC4472 &     8.5 &    0.24 &    0.25 &      12 & 1\\
   NGC4478 &     4.1 &    0.32 &    0.18 &      12 & 1\\
   NGC4489 &     2.6 &    0.24 &   0.09 &      12 & 1\\
   NGC4551 &     5.2 &    0.28 &    0.17 &       7 & 1\\
   NGC4552 &     9.6 &    0.28 &    0.27 &       5 & 1\\
   NGC4564 &     5.9 &    0.44 &     0.2 &       7 & 1\\
   NGC4649 &      11 &    0.30 &     0.3 &       5 & 1\\
   NGC4697 &     8.2 &    0.08 &    0.27 &       5 & 1\\
   NGC4754 &      -- &    $>$0.5 &    0.11 &       4 & 3\\
   NGC4762 &      -- &    $>$0.5 &    0.19 &       4 & 3\\
   NGC4807 &     5.2 &    0.23 &     0.2 &      10 & 1\\
   NGC4816 &     8.3 &    0.20 &    0.25 &      10 & 1\\
   NGC4827 &      10 &    0.18 &    0.24 &      10 & 1\\
   NGC4839 &      15 &    0.07 &    0.25 &      12 & 1\\
   NGC4840 &     6.6 &    0.32 &    0.23 &      10 & 1\\
  NGC4841A &      12 &    0.11 &    0.19 &      10 & 1\\
   NGC4860 &      12 &    0.24 &    0.28 &      10 & 1\\
   NGC4865 &      -- &    $>$0.5 &    0.18 &      10 & 3\\
   NGC4869 &      15 &    0.12 &    0.22 &      10 & 1\\
   NGC4871 &      12 &   -0.02 &    0.22 &      10 & 1\\
   NGC4872 &     2.8 &    0.36 &    0.16 &      10 & 1\\
   NGC4874 &      13 &    0.14 &    0.23 &       3 & 1\\
   NGC4876 &     2.1 &    0.24 &    0.28 &      10 & 1\\
   NGC4883 &      11 &    0.11 &    0.19 &      10 & 1\\
   NGC4884 &      -- &    $>$0.5 &    0.25 &      10 & 3\\
   NGC4895 &      10 &    0.03 &    0.23 &      10 & 1\\
   NGC4896 &      10 &   -0.02 &    0.18 &      10 & 1\\
   NGC4908 &      12 &    0.05 &    0.24 &      10 & 1\\
   NGC4923 &     8.5 &    0.11 &    0.22 &      10 & 1\\
   NGC4926 &      13 &    0.08 &    0.32 &      10 & 1\\
   NGC4931 &     2.8 &    0.30 &    0.14 &      10 & 1\\
   NGC4944 &     2.9 &    0.23 &    0.16 &      10 & 1\\
   NGC4952 &     6.6 &    0.21 &    0.25 &      10 & 1\\
   NGC4957 &     4.9 &    0.32 &    0.19 &      10 & 1\\
   NGC5018 &     1.5 &    0.37 &    0.11 &       9 & 1\\
   NGC5582 &     $>$17 &      -- &    0.24 &       7 & 3\\
   NGC5638 &       7.0 &    0.23 &    0.24 &       5 & 1\\
   NGC5812 &       5.0 &    0.39 &    0.22 &       5 & 1\\
   NGC5813 &     $>$17 &      -- &     0.3 &       5 & 3\\
   NGC5831 &     2.6 &    $>$0.5 &    0.21 &      12 & 1\\
   NGC5846 &      12 &    0.19 &    0.26 &       5 & 1\\
   NGC5866 &     1.8 &    0.35 &   0.10 &       4 & 1\\
   NGC6010 &     4.3 &    0.39 &      -- &       4 & 1\\
   NGC6127 &     9.3 &    0.24 &    0.24 &       5 & 1\\
   NGC6702 &     1.9 &    0.48 &    0.12 &       5 & 1\\
   NGC6703 &     4.1 &    0.33 &    0.19 &       5 & 1\\
   NGC6734 &     4.5 &    0.18 &    0.19 &       9 & 1\\
   NGC6736 &     $>$17 &      -- &    0.24 &       9 & 3\\
   NGC6776 &     3.2 &    0.43 &      -- &       9 & 1\\
   NGC6829 &     4.3 &    0.39 &      -- &       9 & 1\\
   NGC6849 &     $>$17 &      -- &    0.18 &       9 & 3\\

\hline

\end{tabular}
\end{table}

\begin{table}
\begin{tabular}{l c c c c c} 
\hline

Galaxy & Age & [Fe/H] & [Mg/Fe] & Source & Quality\\
\hline

   NGC6958 &      12 &   -0.10 &     0.2 &       9 & 1\\
   NGC7052 &      11 &    0.22 &    0.27 &       5 & 1\\
   NGC7264 &     4.3 &    0.19 &      -- &       5 & 1\\
   NGC7284 &     7.8 &    0.20 &    0.28 &       9 & 1\\
   NGC7332 &     4.5 &    0.24 &      -- &      12 & 1\\
   NGC7396 &     2.6 &    0.39 &      -- &      12 & 1\\
   NGC7454 &     5.2 &   -0.09 &    0.15 &       5 & 1\\
   NGC7562 &      11 &   -0.06 &     0.3 &       5 & 1\\
   NGC7619 &       9 &    0.25 &    0.26 &      12 & 2\\
   NGC7626 &      11.7 &    0.19 &    0.28 &      5 & 1\\
   NGC7703 &     2.6 &    0.29 &      -- &      12 & 1\\
   NGC7768 &     $>$17 &   -0.03 &      -- &       3 & 3\\
   NGC7785 &     8.3 &    0.19 &    0.23 &       5 & 1\\
   NGC7814 &      10 &   -0.10 &      -- &       5 & 1\\
    IC0843 &     $>$17 &      -- &    0.28 &      10 & 3\\
    IC1711 &     6.5 &    0.06 &      -- &      10 & 1\\
    IC1963 &     -- &    $>$0.5 &    0.11 &       8 & 3\\
     IC2006 &       4.0 &    $>$0.5 &    0.19 &       8 & 1\\
    IC3947 &     $>$17 &      -- &    0.33 &      10 & 3\\
    IC4041 &     2.3 &    0.36 &    0.13 &      10 & 1\\
    IC4045 &      14 &    0.07 &    0.23 &      10 & 1\\
    IC4051 &      12 &    0.20 &    0.29 &      10 & 1\\
    IC5011 &      11 &    0.13 &    0.17 &       9 & 1\\
    IC5105 &     $>$17 &    0.08 &    0.21 & 9 & 3\\
   IC5250A &      -- &    $>$0.5 &    0.19 &       9 & 3\\
   IC5250B &     4.3 &    0.33 &     0.2 &       9 & 1\\
    IC5328 &     $>$17 &    0.02 &    0.16 & 9 & 3\\
    IC5358 &      16 &   -0.00 &    0.25 &       9 & 1\\
  IC5364N1 &     2.6 &    $>$0.5 &    0.16 & 9 & 2\\
  IC5364N2 &     $>$17 &      -- &    0.24 &       9 & 3\\
    UGC10043 &     7.3 &  $<$-0.23 &      -- &  9 & 2\\
    UGC11587 &     1.9 &    $>$0.5 &      -- &  9 & 2\\
ESO4710191 &     3.3 &    0.45 &    0.24 &       9 & 1\\
  E107-G04 &     1.8 &    $>$0.5 &    0.14 & 9 & 2\\
  E138-G29 &     9.9 &    0.12 &    0.22 &       9 & 1\\
  E240-G10 &      13 &    0.06 &    0.24 &       9 & 1\\
  E274-G06 &      12 &    0.09 &    0.32 &       9 & 1\\
  E283-G19 &     $>$17 &      -- &    0.38 &       9 & 3\\
  E283-G20 &     1.9 &    0.41 &    0.25 &       9 & 1\\
  E297-G34 &     $>$17 &      -- &    0.15 &       9 & 3\\
  E358-G25 &       -- &   -- &   0.00 &       8 & 3\\
  E358-G50 &       -- &   -- &  0.00 &       8 & 3\\
  E358-G59 &       -- &   -- &    0.14 &       8 & 3\\
  E358-G06 &       -- &   -- &   0.09 &       8 & 3\\
  E359-G02 &       -- &   -- &  -0.04 &       8 & 3\\
  E400-G30 &      -- &      -- &  -0.07 &       9 & 3\\
  E486-G17 &     7.3 &    0.03 &      -- &       9 & 1\\
  E486-G19 &       2.0 &    0.28 &      -- &       9 & 1\\
  E486-G29 &     $>$17 &      -- &      -- &       9 & 3\\
  E507-G45 &     4.2 &    0.45 &    0.23 &       9 & 1\\
  E507-G46 &     $>$17 &      -- &    0.22 &       9 & 3\\
  E538-G10 &      11 &   -0.03 &    0.22 &       9 & 1\\
  E539-G11 &      14 &   -0.24 &    0.28 &       9 & 1\\
  E545-G40 &     $>$17 &      -- &    0.13 &       9 & 3\\

\hline

\end{tabular}

\phantom{a}
\vspace{0.1in}
\noindent
Notes: Sources of age (in Gyrs), metallicity and abundance ratios are: 
Carollo et al. (1993) = 1; 
 Davies et al. (1993) = 2;
 Fisher et al. (1996) = 3; 
 Fisher et al. (1995) = 4; 
 Gonzalez (1993) = 5;
 Goudfrooij et al. (1999) = 6;
 Halliday (1998) = 7;
 Kuntschner (1998) = 8;
 Longhetti et al. (1998) = 9; 
 Mehlert et al. (1997) = 10;
 Vazdekis \& Arimoto (1999) = 11; 
 mean of several = 12. Ages with errors $< \pm 20\%$ and
metallicity errors $< \pm 0.1$ dex have quality index = 1; Errors
$> \pm 20\%$ and $> \pm 0.1$ dex have index = 2; Highly
uncertain values (not used in analysis) have index = 3.

\end{table}

Our final catalogue is given in Table 2. Here we list the galaxy name,
along with our estimate for galaxy age in Gyrs, [Fe/H] metallicity and
[Mg/Fe] abundance. The source of the original line index measurements
are also given; in the case of multiple measurements we have taken the
simple mean. We also list a quality index. Most galaxies have
index = 1 indicating that their age and metallicity estimates are
accurate to $< \pm 20\%$ and $< \pm 0.1$ dex respectively. This
is of course in terms of how their measurement error translates onto
the model grid, and does not include any error associated with a 
systematic shift of the grid. Index = 2 galaxies have larger errors.
About a quarter of the galaxies were found to lie outside
of the Worthey \& Ottaviani (1997) stellar model grid, i.e. usually
suggesting ages greater than 17 Gyrs or metallicities higher than
[Fe/H] = 0.5.  In these cases we give an upper limit. The very high
[Fe/H] galaxies tend to be the most luminous galaxies, and are also
those with super-solar [Mg/Fe] abundances. The galaxies indicating
ages of $>$ 17 Gyrs could be either genuinely old {\it or} suffer from
residual H$\beta$ emission. Thus extreme metal--rich and old galaxies
should be regarded with caution. They are not included in the
subsequent plots or analysis. These, along with galaxies that
don't have [MgFe] line indices available, are assigned index =
3.

\subsection{Comparison with other age estimates}

\begin{table}
\caption{Other age estimates using different methods to the one
utilised for this paper. The sources of these ages are 1 = van Gorkom
et al. (1986); 2 = Brown et al. (2000); 3 = Forbes et al. (1998); 4 =
Whitmore et al. (1997); 5 = Balcells (1997); 6 = Silva \& Bothun
(1998); 7 = Sansom et al. (1998); 8 = Bergvall et al. (1989). }
\centering{
\begin{tabular}{l c c c } 
Galaxy & Our Age & Other Age & Source\\
\hline

NGC 1052 & -- & 0.9 & 1\\
NGC 1700 & 2.3  & 2.5 & 2\\ 
NGC 2865 & $<$1.5 & 1.2 & 3\\
NGC 3156 & -- & 0.9 & 3\\
NGC 3610 & -- & 1.5--6.5 & 4\\ 
NGC 3656 & -- & $\sim$1 & 5\\
NGC 3921 & -- & 0.8 & 3 \\
NGC 5322 & -- & 1--3 & 6\\
NGC 6776 & 3.2 & 2--3 & 7\\
NGC 7252 & -- & 0.7 & 3\\
NGC 7585 & -- & $\sim$1 & A~stars\\
ESO341-IG04 & -- & 0.8 & 8\\

\hline

\end{tabular}}
\end{table}

In Table 3, we have listed age estimates for young, or
proto--ellipticals that we are aware of in the literature. These
estimates come from stellar dynamics, proto--globular clusters,
extended structures etc. For the three galaxies that also have
spectral age estimates from this paper, the agreement is very good.

\begin{figure}
\centerline{\psfig{figure=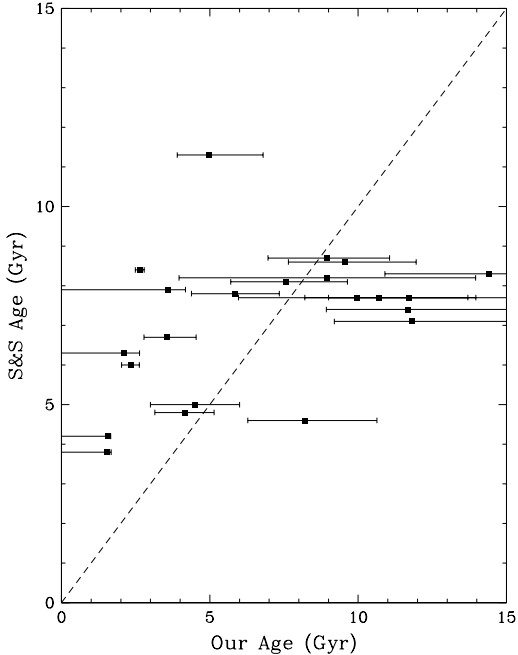,angle=0,width=200pt}}
\caption{ Comparison between ages derived by Schweizer \& Seitzer
(1992) and by us.  The agreement is poor. See text for details.  }
\label{fig:comp_SS}
\end{figure}

\begin{figure}
\centerline{\psfig{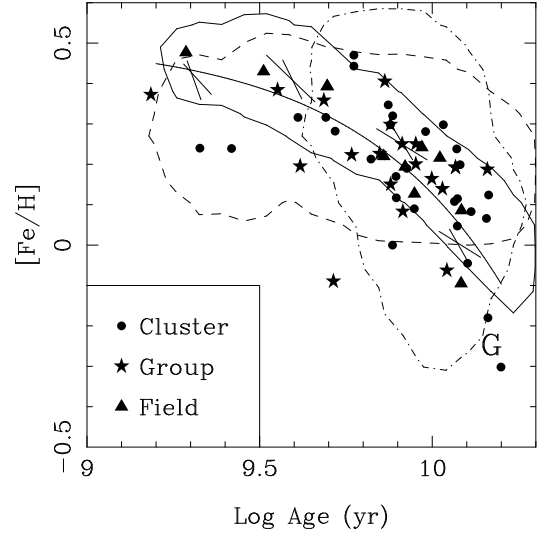}}
\caption{ Metallicity--age relation for all galaxies. Symbols are
coded by environment. Galaxies with younger central components are
more metal--rich. The G symbol in this and subsequent figures shows
the estimated position of the Milky Way (Renzini 1999). The error bars
show how typical errors in H$\beta$ and [MgFe] look in different parts
of the metallicity--age plane. The contours show how the correlated
errors in age and [Fe/H] affect the positions of galaxies in this plot
(see section \protect\ref{sec:corr}). The solid line contours contain
most ($99.56\%$) of the modeled galaxies, whose initial
age-metallicity distribution is known. The dashed line represents a
single metallicity ([Fe/H]$=0.21$) galaxy population. The dot-dash
line shows the distribution of a single age (8.5 Gyrs) population. The
solid line shows the distribution of a population with a linear
relation between age and [Fe/H]. Using a 2d KS test, both
the constant age and constant metallicity models are found to be
inconsistent with the data at greater than 99\% significance
levels. The linear relation is found to be 
acceptable.}
\label{fig:t_Z}
\end{figure}

Another set of age estimates for early type galaxies comes from
Schweizer \& Seitzer (1992). In their paper, they attempted to
quantify the `morphological disturbance' of their sample galaxies, and
via a starburst model, estimate the time since a merger. Depending on
the gas fraction and star formation efficiency, they listed six age
estimates for each galaxy. In Figure \ref{fig:comp_SS} we plot their
``most representative'' age against our age estimate. We note that the
starburst model used by Schweizer \& Seitzer (1992) assumed solar
metallicity. Figure \ref{fig:t_Z} indicates that the young stellar
component is usually metal--rich (i.e. [Fe/H] $\sim$ 0.5 using [MgFe]
as a metallicity indicator).  By assuming solar metallicity we would
expect the Schweizer \& Seitzer ages to be an overestimate.

Examination of Figure \ref{fig:comp_SS} reveals substantial scatter,
with very little formal correlation. We do not believe the Schweizer
\& Seitzer ages to be as reliable as the spectral ages given here but
they do give some indication of whether the galaxy is young or old.

\section{Results and Discussion}
\label{sec:results}

\begin{table*}
\caption{Galaxy properties. Due to its size, the full table is
only available electronically. Morphological type is from the NASA
Extragalactic Database. The environment is given either as Field,
Group or the name of the cluster. The distance is in Mpc, and assumes
H$_o$ = 75. The magnitude, rotation velocity, velocity dispersion,
ellipticity and R (fundamental plane residual) come from Prugniel \&
Simien (1996), Hypercat (Prugniel \& Golev 1999) or RC3. Ellipticities
for S0 galaxies are not listed. } \centering{
\begin{tabular}{l l l c c c c c c} 
Galaxy & Type & Environ & Dist. & M$_B$ & V & $\sigma$ & Ellip & R\\
\hline

   NGC0221 &     cE &  Group &   0.72 & -15.42 &     46 &     76 &     -- &  -0.04 \\
   NGC0224 &     Sb &  Group &   0.72 & -19.94 &    100 &    173 &     -- &  -0.35 \\
   NGC0315 &      E &  Group &    58.9 &  -22.2 &     32 &    299 &   0.31 &  -0.07 \\
   NGC0507 &     S0 &  Group &    67.6 & -21.95 &     -- &    329 &     -- &   0.24 \\
   NGC0522 &     Sc &  Group &    38.5 & -18.99 &     -- &     -- &     -- &     -- \\
   NGC0547 &     E1 &   A194 &    63.4 & -21.84 &     -- &    240 &   0.07 &   0.00 \\
   NGC0584 &     E4 &  Group &    22.2 & -20.42 &    157 &    223 &   0.37 &  -0.19 \\
   NGC0636 &     E3 &  Group &    22.3 & -19.54 &     74 &    166 &   0.14 &  -0.10 \\
   NGC0678 &    SBb &  Group &    39.3 & -19.64 &     -- &     -- &     -- &     -- \\
   NGC0720 &     E5 &  Group &    20.8 & -20.47 &     48 &    244 &   0.37 &   0.05 \\
   NGC0813 &     S0 &  Field &     107 & -21.33 &     -- &     -- &     -- &     -- \\
   NGC0821 &     E6 &  Field &    21.0 & -19.92 &     89 &    207 &   0.37 &   0.15 \\
   NGC0891 &     Sb &  Field &    8.83 & -18.92 &     -- &     -- &   -- &     -- \\
...\\
\hline
\end{tabular}}
\end{table*}

Various properties for the galaxies in our catalogue are listed in
Table 4. This includes the Hubble type from NED\footnote{The NASA/IPAC
Extragalactic Database (NED) is operated by the Jet Propulsion
Laboratory, California Institute of Technology, under contract with
the National Aeronautics and Space Administration.}. Galaxies are
placed in one of three environments, either as a cluster galaxy if in
a known cluster, a group galaxy if in the group catalogue of Garcia
(1993), or in the field if not classified as group or cluster. Galaxy
ellipticity comes from either Davies \etal (1983) or the RC3.  The
other properties are generally taken from Prugniel \& Simien
(1996). The distance includes Virgocentric and Great Attractor
correction terms for H$_o$ = 75 km s$^{-1}$ Mpc$^{-1}$. When a galaxy
was not listed in Prugniel \& Simien, we used Hypercat (Prugniel \&
Golev 1999) and NED. Below we compare the new age, metallicity and
abundance information from Table 2 with other galaxy properties.

\subsection{Galaxy Metallicity -- Age Relation}
\label{sec:TZreln}
In Figure \ref{fig:t_Z} we show the metallicity -- age relation for
the galaxies in our catalogue. The figure also shows model results
showing the distribution of galaxies with given age--metallicity
distributions, taking into account the correlated errors (see section
\ref{sec:corr} and error bars in Figures \ref{fig:t_Z} and
\ref{fig:t_Z_fornax}). The contours show a best fit constant age
distribution (dot dash), a best fit constant [Fe/H] distribution
(dash), and a best fit linear relation between age and [Fe/H] (solid
contour). This linear relation (${\rm [Fe/H]} = 0.51 - 0.38t$, where
the age ($t$) is measured in Gyrs) is shown by the solid line in
Figure \ref{fig:t_Z}.

 There is a trend for the younger galaxies to have more metal--rich
stellar populations. This is as would be expected if the more recent
starbursts are occurring in progressively more enriched gas. The shape
of the metallicity -- age relation is similar to that for local stars
in the disk of our Galaxy (Rana 1991). While there may be some
evidence for the field galaxies preferentially populating the low age,
high metallicity part of the plot, and cluster galaxies tending to be
in the high age, low metallicity portion, there is no strong 
difference in the relation for galaxies in different environments from
this plot. 

We have used a KS test, using models which take into account the
correlated errors (section \ref{sec:corr}) to test what sort of
underlying age--metallicity distribution can best describe the
observed age--metallicity distribution. All three models were
optimised to minimise the KS statistic (i.e. maximise the fit to the
data). The KS test immediately rules out both the single mean age, and
single mean metallicity models to greater than $99\%$.  The linear relation
between age and metallicity (solid line in Figure \ref{fig:t_Z}) has a
probability of being inconsistent with the data of only $88\%$, and
thus can not be ruled out. We thus conclude that only a galaxy
population with a relation between age and metallicity is consistent
with the data. The exact form of the relation though is beyond the
scope of this work. We note that Trager et al. (2000) also favour a
linear relation for the subset of galaxies from Gonzalez (1993) and
Kuntschner (2000). We comment further on this when we discuss
variations of age and metallicity with environment in section
\ref{sec:t_Z_env}




\subsection{Age Variations with Environment} 
\label{sec:t_Z_env}

\begin{figure}
\centerline{\psfig{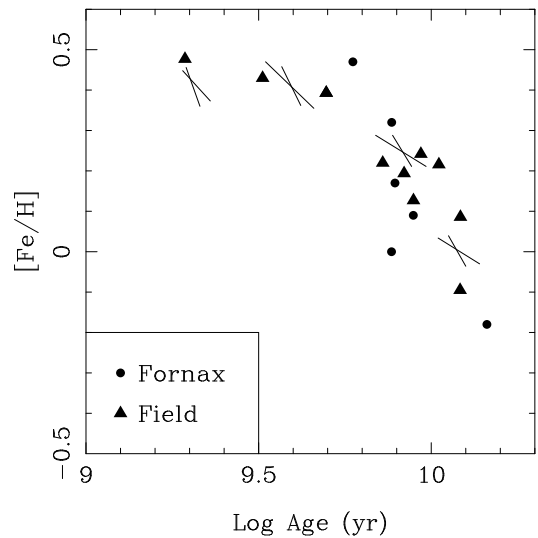}}
\caption{ Metallicity -- age relation for field ellipticals and Fornax
cluster ellipticals. As in Figure \protect\ref{fig:t_Z}, the error
bars show how typical errors in H$\beta$ and [MgFe] look in different
parts of the Metallicity--age plane. Both field and cluster ellipticals
appear to trace out the same metallicity -- age sequence as the
general catalogue (see Figure \protect\ref{fig:t_Z}), i.e. there is not
strong support for the claims that field ellipticals are an age
sequence and cluster ellipticals form a metallicity sequence.  }
\label{fig:t_Z_fornax}
\end{figure}

\begin{figure}
\centerline{\psfig{figure=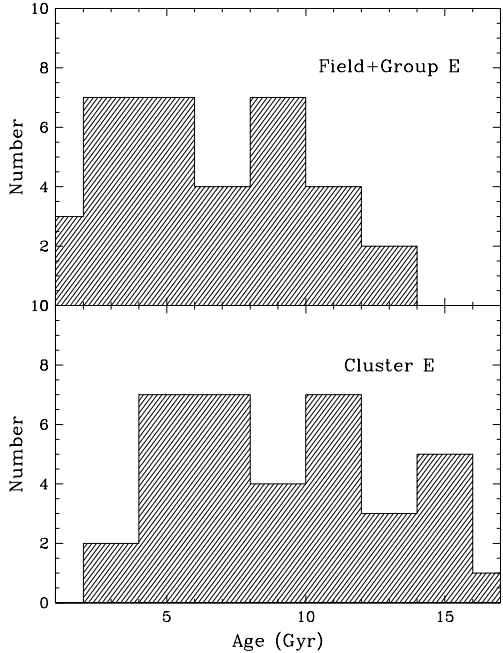,width=200pt}}
\caption{Distribution of ages for elliptical galaxies in clusters
(lower panel) and non-cluster (upper panel). The ellipticals in
non--cluster environments are on average younger by 1.9 Gyrs than their
cluster counterparts.}
\label{fig:all_t_N}
\end{figure}

In his thesis, Gonz\'{a}lez (1993) claimed that field ellipticals
varied greatly in age, at roughly constant metallicity. Kuntschner \&
Davies (1998), on the other hand said that the ellipticals in the
Fornax cluster were essentially a metallicity sequence at constant
age. However, Gonz\'{a}lez included in his sample both S0 and
non--cluster elliptical galaxies. Using only field ellipticals from
our catalogue, many of which come from Gonz\'{a}lez (1993), and the
Fornax ellipticals from Kuntschner \& Davies we reexamine these claims
in Figure \ref{fig:t_Z_fornax}.  This plot shows that both field and
cluster ellipticals are generally consistent with the overall
metallicity -- age relation (see Figure \ref{fig:t_Z}), and that
neither are strictly a sequence in metallicity or age alone (see also
Kuntschner 1998). In fact it would seem that the oldest galaxies span
a range in metallicities, while the younger galaxies all have a high
metallicity.

In Figure \ref{fig:all_t_N} we show the distribution of elliptical
galaxies as a function of environment.  The ellipticals in
non--cluster environments tend to be slightly younger than their
cluster counterparts, however this result is only 98.5\% significant,
and needs to be confirmed with improved data.

This leads to an interesting interpretation of the apparent
Gonz\'{a}lez--Kuntschner discrepancy. While the galaxies from both
studies follow the overall age--metallicity relation shown in Figure
\ref{fig:t_Z}, the Kuntschner ellipticals are
on average older than the Gonz\'{a}lez galaxies, so they span a
range in metallicities, while the younger Gonz\'{a}lez galaxies lie on
a portion of Figure \ref{fig:t_Z} where there are only high
metallicity galaxies.

\subsection{Deviations from the Fundamental Plane}

\begin{figure}
\centerline{\psfig{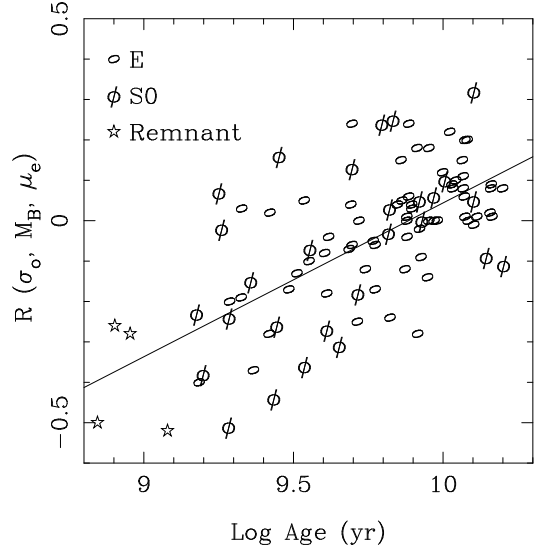}}
\caption{
Residuals from the B band fundamental plane (FP) versus galaxy age. A
similar trend to that first reported by Forbes \etal (1998) is seen. A fit
to the elliptical and S0 galaxies is shown by a solid line, and is
consistent with the location of four $\sim$ 1 Gyr old merger remnants. The
ellipticals show a tighter relation than the S0s. 
}
\label{fig:FPR}
\end{figure}

In an earlier paper (Forbes, Ponman \& Brown 1998) we showed that a
galaxy's position relative to the B band fundamental plane (FP) was
largely due to its age. In particular, young galaxies had negative
residuals and old galaxies positive ones for the FP defined as :\\
R($\sigma_o, M_B, \mu_e) = 2log(\sigma_o) + 0.286M_B + 0.2\mu_e -
3.101$.\\ In that paper, most ages came from spectroscopic estimates
but it was also supplemented by other methods (e.g Schweizer \&
Seitzer 1992). In Figure \ref{fig:FPR} we re-plot the FP residual
diagram using our new galaxy ages (i.e. all spectroscopic). Residuals
from the FP come from Prugniel \& Simien (1996) or calculated from
Hypercat (Prugniel \& Golev 1999). A similar trend, as reported by
Forbes \etal (1998), is seen. A fit to both the elliptical and S0
galaxies gives a slope of $0.36 \pm 0.02$ (bootstrap errors on the
fit). The scatter is considerably reduced if one only considers
elliptical galaxies. The figure also shows the location of four $\sim$
1 Gyr old merger remnants. This relation can be understood in terms of
a centrally located starburst that fades with time (Forbes \etal
1998).

\subsection{Kinematic Trends with Age}

Trager (1997) examined the relationship between galaxy (log) age and
both isophotal shape and internal kinematics for $\sim$40 ellipticals.
He found `L' shaped plots, in the sense that galaxies only occupied
three quadrants of the possible parameter space. Old ellipticals with
disky isophotes and/or isotropic rotators were absence from the plot.
Was this simply due to a small sample or is this an important
evolutionary clue about elliptical galaxies ?

Here we focus on kinematic properties rather than isophotal shape, as
the latter can be strongly influenced by orientation effects. However
we note that isophotal shape (as given by the 4th cosine term) is
correlated with the anisotropy parameter (Kormendy \& Bender 1996), so
any trends with kinematics can probably be extended to isophotal shape
as well.

\begin{figure}
\centerline{\psfig{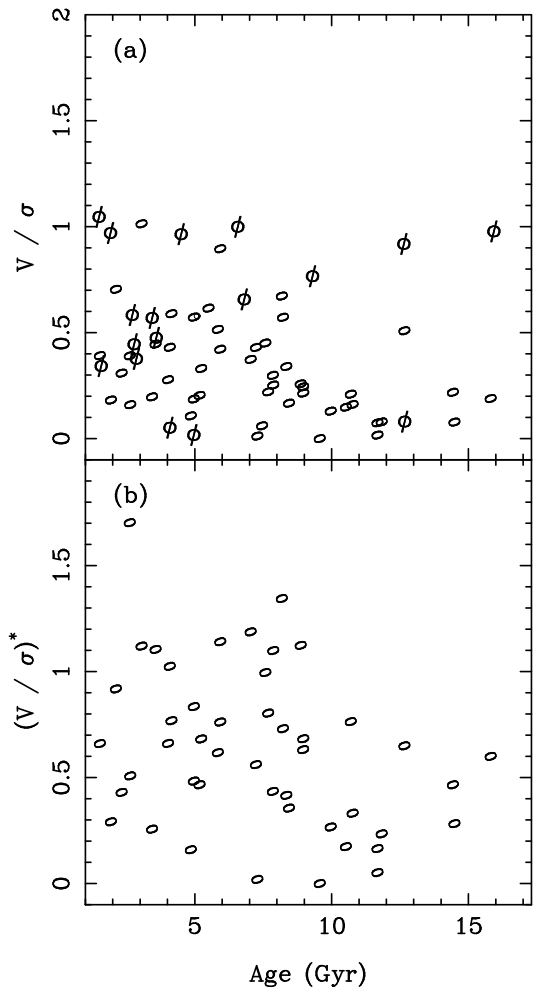}}
\caption{
Rotation properties (maximum rotation velocity relative to velocity
dispersion) versus age. Generally only young ellipticals and S0s have high
${\rm V} / \sigma$ ratios. 
}
\label{fig:t_VoS}
\end{figure}

In Figure \ref{fig:t_VoS}a we plot $V/\sigma$ versus galaxy age.  In
general, the plot shows a tendency for old galaxies to be slow
rotators, while young galaxies can be either slow or rapid
rotators. Furthermore in S0 galaxies rotation dominates over random
motions relative to their elliptical counterparts.  Although the
galaxies do not separate neatly into three quadrants, there is still a
general tendency for fewer old, rapidly rotating galaxies.

Another important kinematic parameter is the anisotropy parameter,
usually denoted as $(V/\sigma)^\ast$.  This is the rotation
parameter normalised by the value for an isotropic oblate spheroid of
a given ellipticity (Davies \etal 1983).  Ellipticals with
$(V/\sigma)^\ast < 0.7$ are said to be anisotropic and those with
$(V/\sigma)^\ast > 0.7$ isotropic, or rotationally flattened.

In Figure \ref{fig:t_VoS}b we plot the anisotropy parameter versus age
for the galaxies classified as elliptical in the RC3. Again there is a
general tendency for fewer old, isotropic galaxies.  Thus our larger
sample generally supports the initial finding of Trager (1997).

If the absence of old, rapidly rotating/isotropic (and hence disky)
ellipticals is real, how can this be explained ?  Selection against
such galaxies in our catalogue seems unlikely, but can't be ruled
out. It is perhaps worthwhile to explore possible physical mechanisms
to explain this effect.We now examine three possibilities. Isotropic
ellipticals \\
{\it i)} did not form in the early Universe.\\
{\it ii)} have been destroyed.\\
{\it iii)} have been transformed into something else.\\

Bender \etal (1992) have suggested that the anisotropy parameter
indicates the relative importance of cold gas to stars in the merger
that created the elliptical. Within this framework, low luminosity
isotropic ellipticals are thought to be the result of gaseous mergers.
As the early Universe is generally more gaseous than today, we would
expect the galaxies formed at early epochs (which are now old) to be
isotropic.  So within this framework, it seems unlikely that isotropic
ellipticals did not form in the early Universe.

For the second option, young isotropic galaxies must be preferentially
destroyed over similarly aged anisotropic ones (e.g via a merger).  It
is again difficult to imagine a scenario in which this is the case.

Transformation of galaxies can take many forms.  Changes in the
dynamical properties of an isolated undisturbed elliptical occurs on
time scales longer than the Hubble time so kinematic transformation via
passive evolution is unlikely.

A potentially important aspect of the kinematic measures discussed
here is that they are luminosity--weighted (as are the age estimates).
The presence of young stars in a disk may dominate the central
spectral indices and hence the galaxy age estimate (de Jong \& Davies
1997). This same disk, with its relatively high $V/\sigma$ ratio, will
also contribute to the overall $V/\sigma$ measurement for the
galaxy. As the disk starburst fades, its relative contribution drops
significantly. So as isotropic ellipticals age they may move not only
to the right in Figure \ref{fig:t_VoS}, but also down as the
elliptical/bulge makes a larger relative contribution to the measured
kinematics.

Some qualitative support for this idea comes from Scorza~\&~Bender
(1995).  They studied the kinematics of 9 ellipticals with disks and
estimated the $(V/\sigma)^\ast$ value for the `bulge' of the
elliptical after subtracting the disk contribution.  The reduction in
the original measured $(V/\sigma)^\ast$ value ranged from 0 to 0.9 in
the sense of making galaxies more anisotropic.

It may also be possible for ellipticals to move from the lower right
of Figure \ref{fig:t_VoS}b to the upper left. If an old anisotropic
galaxy accretes a small gaseous galaxy, which leads to the formation
of a gaseous disk and associated star formation, then this could have
the effect of resetting the galaxy age while making it appear
isotropic as well.

Disky, isotropic ellipticals tend also to have high ellipticities and
reveal power--law inner surface brightness profiles (e.g. Kormendy \&
Bender 1996).  We examined the distribution of ages for galaxies with
`power--law' versus `core' surface brightness profiles, and found a
weak trend for power--law galaxies to be younger on average, with very
few old power--law ellipticals.  This is in the same sense as expected
from Figure \ref{fig:t_VoS}.  However there were insufficient numbers
of galaxies in our sample to produce statistically significant
results.


\subsection{Predictions from Hierarchical Clustering and Merging Models}

In a hierarchical Universe galaxies are built up from the merging of
smaller subunits. Starting from an extended Press--Schechter theory,
various groups have developed semi--analytical models to describe
galaxy formation via hierarchical clustering and merging (HCM;
Kauffmann \etal 1993; Cole \etal 1994; Somerville \& Primack
1999). Although some differences exist between the different
formulations, they have some basic characteristics of galaxy formation
in common. For example, HCM requires that massive ellipticals are
younger than less massive ones. Environment is a key property in
determining galaxy evolution under HCM.  Ellipticals in low density
environments (i.e. the field) have a more complex and extended star
formation history making them on average younger than their cluster
counterparts.

In the following we compare the results from our catalogue with the
HCM predictions of Kauffmann and co--workers (e.g. Kauffmann \&
Charlot 1998; Thomas \& Kauffmann 1999). Although our catalogue of
galaxies is in no sense statistically complete, it could be considered
a random sample and therefore fairly representative of nearby
galaxies.  Our ages are measuring the last major episode of star
formation, which is presumably an indicator of the time since {\it
assembly} of the galaxy in a dynamical sense, and not the mean age of
the stars in a galaxy.  We attempt to make quantitative comparisons
where possible, however one should bear in mind differences between
`observers' and `theorists' definitions. These are briefly outlined
below in the four predictions:\\

\begin{figure}
\centerline{\psfig{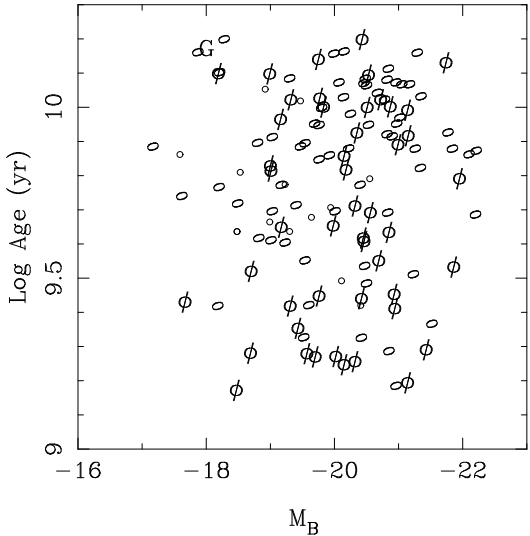}}
\caption{ Luminosity versus age for early type galaxies. There is no
strong trend for either ellipticals or S0 galaxies. The symbols used
are the same as for Figure \protect\ref{fig:Worthey_gals} }
\label{fig:MB_t}
\end{figure}

\noindent
{\it i) Large ellipticals are younger than small ellipticals.}\\ A
basic requirement of all HCM models is that more massive ellipticals
are formed more recently than less massive ones and hence should be
younger on average. Based on a sample of $\sim$40 galaxies, Faber
\etal (1995) suggested that large ellipticals were {\it older} on
average than small ellipticals. However, in our previous paper (Forbes
\& Ponman 1999), our sample of 88 galaxies did not show a strong
luminosity -- age trend. In Figure \ref{fig:MB_t} we show the
distribution of total B band luminosity with galaxy age for the
elliptical and S0 galaxies in our catalogue. We find no statistically
significant correlation in our sample. This result supports those of
colour--magnitude studies (e.g Terlevich et. al. 1999), but it is also
possible that environment effects are obscuring any trend; we examine
this next. \\

\noindent
{\it ii) Field ellipticals are younger than cluster ellipticals.} \\
In the models, galaxy `environment' is defined in terms of the
circular velocity (V$_C$) of the dark matter halos, with `field'
ellipticals having V$_C$ $<$ 600 km s$^{-1}$ and `cluster' galaxies
halos having V$_C \sim 1000$ km s$^{-1}$ (Kauffmann \& Charlot 1998).
The V--luminosity weighted age difference between these cluster and field
ellipticals ranges from about zero for small ellipticals to 1 Gyr for
the most massive ellipticals. HCM has received some support from the
work of Bernardi \etal (1998) who claim differences in the Mg --
$\sigma$ relation for cluster and field ellipticals indicating an age
difference of 1.2 $\pm$ 0.35 Gyrs in the correct sense. In our
catalogue, we have classified galaxies into known clusters, groups (if
in the list of Garcia 1993) and field (essentially non--group and
non--cluster galaxies).  
In Figure \ref{fig:all_t_N} we show the age
distribution for field and group galaxies compared to their cluster
counterparts. 
We find that the mean age of elliptical
galaxies in clusters is 9.0 $\pm$ 0.7 Gyrs (error on the mean),
compared to 7.1 $\pm$ 0.6 for field and group ellipticals. The age difference of
1.9 Gyrs is similar to that claimed by Bernardi \etal (1998), although
the small numbers of galaxies means that our result is not
statistically significant. \\

\begin{figure}
\centerline{\psfig{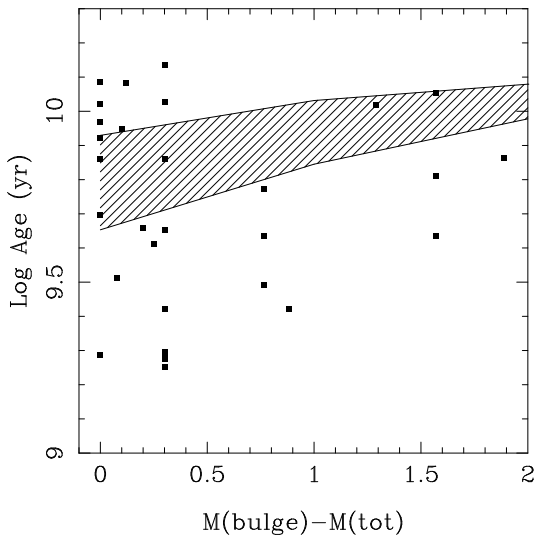}}
\caption{
Distribution of galaxy age with Hubble type for field galaxies. 
Here Hubble type is
represented by the magnitude of the bulge component minus the total galaxy
magnitude (e.g. ellipticals are 0 on this scale, Sb spirals are 1.57). 
There is little or no trend in the data. The solid lines region the region
occupied by the HCM models of Kauffmann (1996) for a V band luminosity
weighted galaxy age. The data do not follow the HCM trend particularly
well, and reveal a population of old field ellipticals that are not 
present in the models. 
}
\label{fig:BTR_t}
\end{figure}

\noindent
{\it iii) Large bulge-to-disk galaxies are younger than small
bulge-to-disk galaxies.}\\ According to HCM, after an elliptical
galaxy forms from the merger of smaller disk systems, subsequent
cooling of the gas in the halo may `accrete' onto the elliptical
forming a bulge and disk system. Thus HCM predicts that spirals can be
transformed into ellipticals and back into spirals again. Furthermore
spirals with large bulge-to-disk ratios (e.g. Sa) have had less time
to develop a disk, and are hence younger than late type spirals (e.g.
Sc).  Kauffmann (1996) shows that for field galaxies, the V band luminosity
weighted age varies from $\sim$6.5 to 12.5 Gyrs for M(bulge) -- M(tot),
where M(bulge) is the B band bulge magnitude and M(tot) the total B
band magnitude of the galaxy.  Figure \ref{fig:BTR_t} shows the galaxy
age versus M(bulge) - M(tot) for field galaxies in our catalogue. We
have translated the galaxy T type (from RC3) into M(bulge) - M(tot)
based on the study of Simien \& de Vaucouleurs (1986). We also show
the HCM predictions of Kauffmann (1986).  The limited data for late
type galaxies (M(bulge) -- M(tot) $> 0.5$) follows a similar overall
trend to the models, but lie outside the range of ages
predicted. 

Most of the spiral galaxies in Figure \ref{fig:BTR_t} come from the
study of Goudfrooij, Gorgas \& Jablonka (1999). They were careful to
reduce the disk light emission as much as possible, but if some
residual disk light does remain in the slit (placed along the
minor axis) then this may bias the late type spirals to younger
ages. They noted that the bulges of spirals have similar ages to
ellipticals.

It is worth mentioning an alternative bulge formation mechanism based
on secular growth from disk material (e.g. Pfenniger \& Norman
1990). In contrast to the `bulge first' formation of HCM, this
scenario has bulges formed after the disk, i.e. `disk first'. So we
would expect early type spirals to be older than late type spirals.
Trager, Dalcanton \& Weiner (1998) have presented line indices for two
early type and two late type spirals. They conclude that the bulges of
the two early type spirals are indeed {\it older} than the late type
ones.  Our data suggest the opposite trend; clearly more data are
needed to resolve this issue.\\

\noindent
{\it iv) Large, field ellipticals have lower [Mg/Fe] ratios than
small, cluster ellipticals.}\\ It is now fairly well established that
massive ellipticals have an enhancement in $\alpha$ elements,
e.g. [Mg/Fe] ratios that are super-solar by about 0.3
dex. (e.g. Peletier 1989; Worthey, Faber \& Gonz\'{a}lez 1992; Davies
\etal 1993; Carollo \etal 1993).  These ratios probably indicate that
massive ellipticals have short star formation timescales and/or an IMF
that is skewed towards high mass stars. Such element abundances
provide an important probe of galaxy formation and evolution.

\begin{figure}
\centerline{\psfig{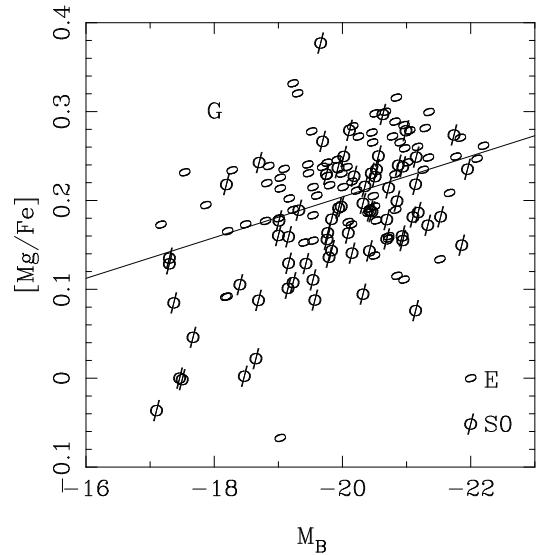}}
\caption{ Abundance versus luminosity. The plot shows that larger
galaxies tend to be over abundant in Mg relative to Fe, with [Mg/Fe]
$\sim$ 0.3 for the most massive galaxies. The symbols are as for
Figure \protect\ref{fig:Worthey_gals} }
\label{fig:MB_overab}
\end{figure}

The incorporation of chemical evolution into the HCM semi analytical
models, and the predictions for [Mg/Fe] ratios, are described in
Thomas \& Kauffmann (1999).  They note that the range of predicted
[Mg/Fe] does not yet match the observations. So rather than make
direct comparisons, we will simply explore [Mg/Fe] trends. The first
trend we compare is that of [Mg/Fe] with luminosity for elliptical
galaxies. The HCM model predicts a general {\it decrease} in the
average [Mg/Fe] ratio for more luminous ellipticals.  As noted above,
the observational data show a clear trend for {\it increasing} [Mg/Fe]
in large ellipticals (see Figure \ref{fig:MB_overab}). This disagreement
between the theory and observation is noted by Thomas \& Kauffmann,
and they go on to suggest that a flatter IMF will help raise the
predicted [Mg/Fe] ratios.  As an aside, the bulges of spirals also
reveal a similar [Mg/Fe] trend with luminosity (Jablonka \etal
1996). It is difficult to explain this trend and understand the
$\alpha$ element enhancement if bulges are being built up slowly
(i.e. on timescales longer than those associated with SN Ia
explosions).

\begin{figure}
\centerline{\psfig{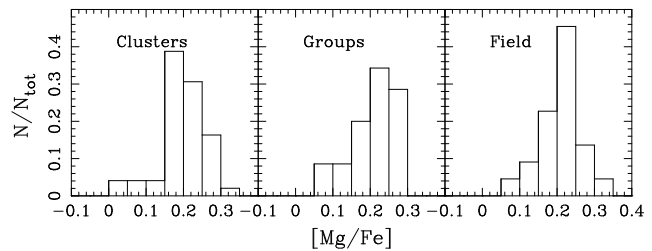}}
\caption{ Distribution of [Mg/Fe] for ellipticals and S0 galaxies in
different environments. The HCM models show a strong preference for
cluster galaxies to have lower [Mg/Fe] ratios than field galaxies,
which is not seen in the data (although clearly the data suffer from
low number statistics).  }
\label{fig:overab_N}
\end{figure}

A second prediction of the model is that field galaxies should exhibit
lower [Mg/Fe] ratios on average than their cluster counterparts.  We
do not find this to be the case (see Figure
\ref{fig:overab_N}).  Thomas \& Kauffmann (1999) do state that their
models under predict the number of galaxies with high [Mg/Fe] compared
to observations, however the shape of the distributions shown
in Figure \ref{fig:overab_N} do not match the distributions of [Mg/Fe]
for ellipticals and bulges predicted by the models, which have a much
flatter distribution, skewed towards lower values of [Mg/Fe]. The few
galaxies in the models which do have [Mg/Fe] ratios comparable with
observations, formed in the first 1--2 Gyrs. 

It is also important to note that massive ellipticals have [Mg/Fe]
over abundant {\it at all radii} not just at the galaxy centre (Davies
\etal 1993; Kuntschner 1998).  This indicates that the exact nature of
the star formation occurred over the whole galaxy. This combined with
the very short star formation time scale implied (i.e. less than the
SN Ia time scale of $\sim1$ Gyr) favours a dissipational collapse
and/or gaseous mergers at early epochs for these galaxies. Lower
luminosity ellipticals, with near solar [Mg/Fe], allow for more
extended star formation, as might be associated with a merger.

\section{Conclusions and Future Work}

Using high quality spectral line index data from the literature we have derived
ages, metallicities and abundance ratios for about 150
galaxies. These, mostly early-type galaxies, cover a range of
luminosities and environments. We confirm previous findings that the
scatter in the elliptical galaxy fundamental plane depends on the
galaxy age. Our data support some predictions of hierarchical galaxy
formation (i.e. that field ellipticals are younger than their cluster
counterparts), but are at odds with others (i.e. massive ellipticals
are not obviously younger than small ellipticals as predicted). We
also find an interesting absence of old, rapidly rotating
galaxies in our sample. 

An outstanding issue in contemporary galaxy studies concerns the apparent 
dichotomy in elliptical galaxy properties. The dichotomy occurs at about 
M$_B$ = --20.5 (although it overlaps in luminosity), with low luminosity 
ellipticals tending to have disky isophotes, power-law surface 
brightness profiles and be isotropic rotators. As such they have much in 
common with S0 galaxies (Kormendy \& Bender 1996). 
Various aspects of the dichotomy debate are presented by van den Bergh 
(1998). He suggests that low luminosity ellipticals may represent the 
product of a dissipational collapse. High luminosity ellipticals, on the 
other hand, formed less dissipatively from stellar mergers according to 
Bender, Burstein \& Faber (1992). However the super solar [Mg/Fe] ratios 
of such galaxies probably indicates a rapid star formation timescale 
within the merging subunits. 

Our results for trends with galaxy age lend some support to this
view. The galaxies with `dissipative features' (e.g. high V/$\sigma$,
high ellipticity, power-law profiles) tend to be young. It is likely
the young stellar population is associated with a stellar disk (as
found by de Jong \& Davies 1997) and a disky E/S0 galaxy reveals
other the dissipative features. As this disk starburst fades with
time, the old stellar population contributes more to the line indices
(and hence the age) but also to the kinematics.

Although we have assembled a large number of relative ages and metallicities
for local galaxies, there are some issues with this dataset which
prevented us from properly testing many of the predictions of
HCM. The main points which need to be addressed in future studies
include:

\noindent{\it i)} Sparsity of field and group galaxies in the sample. Many
predictions of HCM compare the properties of field, cluster and group
galaxies, however the field and group galaxy data available to date is
of lower quality than the cluster data.

\noindent{\it ii)} Nebular emission (see section \ref{sec:emission}) could
still be affecting the ages and metallicities we derive for some of
the galaxies in our sample. This could be minimised in future studies
by using higher order Balmer line indices, such as H$\gamma$ and
H$\delta$.

\noindent{\it iii)} The sample definition in this paper has by
necessity been very simple. We have used all data which satisfied our
signal-to-noise criterion. 
Future work should concentrate on creating a far more uniformly selected
sample, which can be regarded as either complete or
representative for a particular galaxy population.\\

\noindent{\bf Acknowledgements}\\ 
We thank H. Kuntschner, T. Ponman, S. Trager, A. Smith, K. Masters for
help and useful discussions. We also thank D. Mehlert, C. Halliday,
Longhetti, P. Goudfrooij for supplying us with electronic versions of
their data.\\

This research has made use of the NASA/IPAC Extragalactic Database
(NED) which is operated by the Jet Propulsion Laboratory, California
Institute of Technology, under contract with the National Aeronautics
and Space Administration.

\vskip 1cm

\noindent{\bf References}

\noindent
Baugh C. M., Cole S., Frenk C. S., Lacey C. G., 1998, ApJ, 498, 504\\
Bender R. , Burstein D., Faber S. M., 1992, ApJ, 399, 462\\
Bernardi M., et al., 1998, ApJ, 508, L143\\
Bower R.~G., Lucey J.~R., Ellis R.~S., 1992, MNRAS, 254, 601\\
Brown R. J. N., Forbes D. A., Kissler--Patig M., Brodie J. P., 2000, 
MNRAS, 317, 406 \\
Bruzual G., Charlot S., 1993, ApJ, 405, 538\\
Burstein D., Faber S. M., Gaskell C. M. , Krumm N., 1984, ApJ, 287, 586\\
Buzzoni A., Gariboldi G., Mantegazza L.,  1992, AJ, 103, 1814\\
Buzzoni A., Mantegazza L., Gariboldi G., 1994, AJ, 107, 513\\
Caldwell N., Rose J.~A., Sharples R.~M., Ellis R.~S., Bower R.~G., 1993, AJ, 106, 473\\
Carollo C. M., Danziger I. J., Buson L., 1993, MNRAS, 265, 553\\
Cole S., Aragon-Salamanca A., Frenk C.~S., Navarro J.~F., Zepf
S.~E., 1994, MNRAS, 271, 781\\
Davies R.~L., Illingworth G., 1983, ApJ, 266, 516\\
Davies R.~L., Sadler E.~M., Peletier R.~F., 1993, MNRAS, 262, 650\\
Ellis R.~S., \etal, 1997, ApJ, 483, 582\\
Faber S.~M., Jackson R.~E., 1976, ApJ, 204, 668\\
Faber S.~M., Trager S., Gonz\'{a}lez J., Worthey G., 1995, in Stellar
Populations, eds P.~C. van der Kruit and G. Gilmore (Dordrecht: Kluwer), 249\\
Fisher D., Franx M., Illingworth G.~D., 1995, ApJ, 448, 119\\
Fisher D., Franx M., Illingworth G.~D., 1996, ApJ, 459, 110\\
Forbes D.~A., Ponman T.~J., Brown R.~J.~N., 1998, ApJ, 508, L43\\
Fritze-V. Alvensleben U., Burkert A., 1995, A\&A, 300, 58\\
Garcia A.~M., 1993, A\&AS, 100, 47\\
Gonz\'{a}lez J.~J., 1993, Ph.D Thesis, University of California, Santa Cruz\\
Gorgas J., Efstathiou G., Salamanca A.~A., 1990, MNRAS, 245, 217\\
Goudfrooij P., Emsellem, E., 1996, A\&A, 306, L45\\
Goudfrooij P., Gorgas J., Jablonka P., 1999, Ap\&SS, 269, 109\\
Governato F., Gardner J. P., Stadel J., Quinn T., Lake G., 1999, AJ,
117, 1651\\
Halliday C., 1998, Ph.D Thesis, University of Durham, UK\\
Jablonka P., Martin P., Arimoto N., 1996, AJ, 112, 1415\\
Jones L. A., 1997, PhD Thesis, Univ, of North Carolina, Chapel Hill\\
de Jong R.~S., Davies R.~L., 1997, MNRAS, 285, L1\\
J$\o$rgensen I.,  1997, MNRAS, 288, 161\\
J$\o$rgensen I., 1999, MNRAS, 306, 607\\
Kauffmann G., White S. D. M., Guiderdoni B., 1993, MNRAS, 264, 201\\
Kauffmann G.,  1996, MNRAS, 281, 487\\
Kauffmann G., Charlot S., 1998, MNRAS, 294, 705\\ 
Kodama T., Arimoto N., Barger A. J., Aragon-Salamanca A., 1998, A\&A,
334, 99\\
Kormendy J., Bender R., 1996, ApJ, 464, L119\\
Kuntschner H., 1998, PhD Thesis, Durham\\
Kuntschner H., 2000, MNRAS, 315, 184\\
Longhetti M., Rampazzo R., Bressan A., Chiosi C., 1998, A\&AS, 130, 251\\
Maraston C., Greggio L., Thomas D., 2001, Ap\&SS, 276, 893\\
Matteucci F., Ponzone R., Gibson B. K., 1998, A\&A, 335, 855\\
Mehlert D., Bender R., Saglia R. P., Wegner G., 1997, in Mazure, A.,
Casoli F., Durret F. , Gerbal D., eds, Coma Berenices: A New Vision of
an Old Cluster.  Word Scientific Publishing Co Pte Ltd, p. 107\\
O'Connell R. W., 1976, ApJ, 206, 370\\
Osterbrock D. E., "Astrophysics of Gaseous Nebulae and Active
Galactic Nuclei", University Science Books, NY 10012, 1989.\\
Peletier R., 1989, PhD Thesis, Groningen\\
Pfenniger D., Norman C., 1990, ApJ, 363, 391\\
Prugniel P., Golev V., 1999, in  Carral P., Cepa J., eds, Star Formation in Early Type Galaxies. ASP Conference Series 163, p. 296\\
Prugniel P., Simien F., 1996, A\&A, 309, 749\\
Rana M. C., 1991, ARAA, 29, 129\\
Renzini A., 1999, in Carollo C.M., Ferguson H.C., Wyse R.F.G., eds, Origin of Bulges. Cambridge University Press, Cambridge. p. 9\\
Schweizer F., Seitzer P., 1992, 104, 1039\\
Scorza C., Bender R., 1995, A\&A, 293, 20\\
Simien F., de Vaucouleurs G., 1986, ApJ, 302, 564\\
Somerville R. S., Primack J. R., 1999, MNRAS, 310, 1087\\ 
Terlevich A. I., Kuntschner H. , Bower R. G., Caldwell N., Sharples R. M., 1999, MNRAS, 310, 445 \\
Thomas D., Kauffmann G., 1999, in Hubeny I., Heap S., Cornett R., eds, Spectrophotometric Dating of Stars and Galaxies. ASP Conference Proceedings, Vol. 192, p. 261\\
Trager S. C., 1997, Ph.D Thesis, University of California, Santa Cruz\\
Trager S. C., Worthey G., Faber S. M., Burstein D., Gonz\'alez J. J., 
1998, ApJS, 116, 1\\
Trager, S.~C., Faber, S.~M., Worthey, G., Gonz\'alez, J.~J., 2000, AJ, 120, 165\\
van den Bergh S.,  1998, ``Galaxy morphology and classification'' Cambridge University Press.\\  
van Dokkum P. G., Franx M., Kelson D. D., Illingworth G. D., 1998,
ApJ, 504, 17\\
Vazdekis A., 1996, PhD Thesis, Univ. of La Laguna, Spain\\
Vazdekis A., Casuso E., Peletier R. F.,Beckman J. E., 1996, ApJS,
106, 307\\
Vazdekis A., Peletier R. F., Beckman J. E., Casuso E., 
1997, ApJS, 111, 203\\
Vazdekis A., Arimoto N., 1999, ApJ, 525, 144\\
Whitmore B. C., Miller B. W., Schweizer F., Fall S. M., 1997, AJ,
114, 1797\\ 
McWilliam A., 1997, ARAA, 35, 503 \\
Worthey G., Faber S. M., Gonz\'{a}lez J. J., 1992, ApJ, 398, 69\\
Worthey G., 1994, ApJS, 95, 107\\
Worthey G., Ottaviani D. L., 1997, ApJS, 111, 377\\
Zabludoff A. I., \etal, 1996, ApJ, 466, 104\\

\label{lastpage}

\end{document}